\documentclass[10pt,dvips]{article}

\usepackage{url}
\usepackage{cite}
\usepackage{feynmp}
\usepackage{amsmath}
\usepackage{amssymb}
\usepackage{bbold}
\usepackage{axodraw}
\usepackage{slashbox}
\usepackage{rotating}
\usepackage{axodraw}
\usepackage{epsfig}
\usepackage{color}
\usepackage{hhline}
\usepackage{graphicx}

\renewcommand{\thefootnote}{\fnsymbol{footnote}}

\begin{document}


\newcommand{\eeHZ}{$e^+e^- \to ZH$}
\newcommand{\epem}{$e^+e^-$}


\vskip 2.cm

\begin{center}
{\large \bf General Spin Analysis from Angular Correlations in Two-Body Decays}\\[1.5cm]
{Seong Youl Choi\footnote{choisy7@gmail.com},\,
 Jae Hoon Jeong\footnote{jeong229@jbnu.ac.kr},\, and
 Ji Ho Song\footnote{sonjiho0@nate.com}}\\[0.5cm]
{\it  Department of Physics and RIPC, Chonbuk National University,
      Jeonju 54896, Korea}
\end{center}

\begin{center}
(\,\today\,)
\end{center}

\vskip 1.cm

\begin{abstract}
\noindent Determining the spin of any new particle and measuring its
couplings to other particles and/or itself are crucial in reconstructing the
structure of any quantum field theory containing the particle. A general
helicity formalism is employed to describe the polarization of the particle
$Y$ in a two-body decay $X_2\to Y X_1$ with polarized $X_2$ for the purpose
of diagnosing the dynamical properties of three involved particles and for
determining their spins altogether. We perform a general and comprehensive
analytic analysis with our special focus on grasping fully how to connect the
decay helicity amplitudes and decay distributions in the $X_2$ rest frame and
those in a laboratory frame with $X_2$ moving with a non-zero velocity
through Wick helicity rotation on helicity states and amplitudes. This
theoretical framework is demonstrated in a detailed illustrative manner with
the Standard Model (SM) processes, the sequential process $e^-e^+\to Z\to
\tau^-\tau^+$ followed by $\tau^-\to \rho^-\nu_\tau\to (\pi^-\pi^0)\nu_\tau$
and the sequential process $e^-e^+\to t\bar{t}$ followed by $t\to W^+ b \to
(\ell^+\nu_\ell)b$--, and one non-standard decay process of a new vectorlike
heavy top quark, $T\to Z t$, followed by $Z\to \ell^-\ell^+$. All the useful
formulas directly applicable to any combinations of spins and any types of
couplings in the two-body decay $X_2\to Y X_1$ followed by suitable $Y$
two-body decays processes are collected and
described in detail. \\[2mm]
\end{abstract}

\vskip 1.cm



\setcounter{footnote}{0}
\renewcommand{\thefootnote}{\arabic{footnote}}

\section{Introduction}
\label{sec:introduction}

Along with mass, spin is a basic invariant property that every elementary
particle and any isolated object must possess in the four-dimensional spacetime
with Lorentz invariance \cite{Wigner:1939cj}. Questions and answers about the
spin dependence of reactions therefore have played an essential role in probing
the underlying theoretical structures very deeply and therefore establishing
the SM of electroweak and strong interactions in elementary particle
physics up to now \cite{Bourrely:1980mr,Bourrely:1987gp,Leader:2001gr}.\\

On the high energy frontier, equipped with the Large Hadron Collider (LHC)
\cite{Evans:2008zzb}, we are now probing the electroweak (EW) scale ($v=246$
GeV) and beyond intensively and extensively after having established the SM by
the decisive discovery of a Higgs boson \cite{Aad:2012tfa,Chatrchyan:2012xdj}
followed by very precise measurements of its mass and couplings
\cite{Aad:2015zhl,Khachatryan:2016vau} and model-independent determinations of
its spinless nature
\cite{Choi:2002jk,Gao:2010qx,DeRujula:2010ys,Bolognesi:2012mm,Aad:2015mxa,Khachatryan:2014kca}.
The true theory for the origin and stability of the EW scale
\cite{Weinberg:1976,Weinberg:1979bn,Susskind:1979,tHooft:1980} beyond the SM is
highly expected to be
revealed with a huge amount of accumulated data.\\

One generic prediction in most of new models is the presence of new particles
partnered with some or all of the SM particles. For instance, every SM particle
in low-energy supersymmetry (SUSY)
\cite{Nilles,Haber:1984rc,Chung:2003fi,Drees:2004jm,Binetruy:2006ad,Wess:1992cp}
has a heavier partner whose spin differs by 1/2 in units of $\hbar$.
Alternatively, in universal extra dimension (UED) models
\cite{Appelquist:2000nn,Kakuda:2013kba}, each SM particle is paired with a
tower of Kaluza-Klein (KK) excitations with identical spin. Thus,
model-independent determinations and detailed measurements of the spins and
dynamical structures are
crucial in discriminating among new scenarios. \\

In the present work a general theoretical framework is presented for describing
the spin and polar-angle correlations\footnote{Generally, azimuthal-angle
correlations can be included in the analysis as well, but they involve quantum
interference among the states with different helicities and require by far more
complicated kinematic reconstructions \cite{Boudjema:2009fz}. For the sake of
simple and straightforward kinematical analyses, we do not consider them here,
postponing the analysis involving azimuthal-angle correlations as our later
project.} in the two-stage two-body decays of a polarized state $X_2$ of spin
$j_2$ and mass $m_2$ into two on-shell states, $Y$ of spin $j$ and mass $m$ and
$X_1$ of spin $j_1$ and mass $m_1$
\begin{eqnarray}
X_2[j_2, m_2]\,\,\to\,\, Y[j, m] + X_1[j_1, m_1]\,,
\label{eq:x-2_y_x-1_decay}
\end{eqnarray}
followed by a two-body decay of the particle $Y$ into a particle $a$ of spin
$j_a$ and mass $m_a$ and a particle $b$ of spin $j_2$ and mass $m_b$
\begin{eqnarray}
Y[j,m]\,\,\to\,\, a[j_a, m_a] + b[j_b, m_b]\,,
\label{eq:y_a_b_decay}
\end{eqnarray}
where at least the momentum of the particle $a$ is assumed to be measurable
event by event.\footnote{In principle any multi-body decay modes of the
particle $Y$ can be considered for extracting the information on $Y$
polarization.} For a non-zero $j_2$, the $X_2$ particle is produced generally
in a polarized state in its production processes, in particular, if the
interactions are parity-violating, and the polarization of the particle $Y$
with a non-zero spin $j$ can be extracted (partially) through the angular
distributions in its sequential decays. If the branching fractions are sizable,
then the sequential two-stage decays can provide us with a powerful tool not
only for examining the properties of the involved particles but also for
determining their spins altogether, as will be demonstrated with specific
examples in
the following.\\

When the rest frame of the decaying particle is hardly reconstructible as in
$pp$ collisions at the LHC, the direct spin measurements are performed
conventionally through a set of Lorentz-invariant masses constructed in
sufficiently long decay chains \cite{Barr:2004ze,Barr:2005dz,Smillie:2005ar}.
Such spin-determination methods tend to rely heavily on a number of final state
spins and involved (chiral) couplings \cite{Wang:2006hk,Wang:2008sw}. In this
work, we will demonstrate with several specific examples that the polar-angle
correlations of the $Y$ in the rest frame of $X_2$ ($X_2$RF) and one of the $Y$
decay products in the $Y$ rest frame (assumed to be reconstructed at least
partially) also enable us to determine the spins and underlying dynamics
decisively and clearly.\\

If the four-momentum of the particle $Y$ or one of its decay products can be
determined event by event even though the momentum of the decaying parent
particle $X_2$ is not reconstructed, one natural reference axis for describing
the $Y$ polarization is nothing but the $Y$ flight direction {\it in the
laboratory frame (LAB)}, to be called the detection axis in the following. In
this situation, the most natural experimental observable for $Y$ decays is then
the polar-angle as well as azimuthal-angle distribution of one of the $Y$ decay
products in the $Y$ rest frame {\it boosted back directly along the $Y$
momentum direction in the LAB}.\\

Certainly, the most convenient reference system for describing the dynamics of
the two-body decay $X_2\to Y X_1$ {\it analytically} without any kinematical
complications caused by boosts or rotations is the {\it $X_2$RF}, (often
difficult or sometimes impossible to reconstruct event by event). As a result,
there exists a subtle mismatch between the transparent theoretical description
in the {\it $X_2$RF} and the direct experimental determination of spins and
dynamical properties in the {\it LAB}. As worked out in detail later, the
quantum state and polarization of the particle $Y$ in the LAB are related to
those in the $X_2$RF by several well-established kinematical functions which
fully encode the impact of the {\it Wick helicity rotation}
\cite{Wigner_rotation}, (closely related but {\it not identical} to the Wigner
rotation \cite{Wigner:1939cj}) that is induced from two
consecutive non-parallel Lorentz transformations.\\

The polarization parameters of $Y$ in the $X_2$RF are given {\it simply} by
dynamical parameters such as spins, couplings, mixing matrices and masses of
the particles involved in the two-body decay. On the contrary, the polarization
parameters of $Y$ in the LAB are connected {\it directly} to the sequential
decay(s) of $Y$ so that they can be measured and determined directly in
experiments as they often do not require the full kinematic reconstruction of
the entire event
chain. \\

One transparent path for connecting the values of polarization parameters
measured experimentally in the LAB with the dynamical theory parameters encoded
in the $X_2$ decay amplitudes is provided by the helicity formalism
\cite{Jacob:1959at,Wick:1962zz,Chung:1971ri}, allowing us to deal with massless
and massive particles on an equal footing . Without any specific assumptions on
particle spins and masses, we provide a general spin analysis for predicting
the LAB values of the polarization parameters and comparing them directly
through angular correlations. In order to cover the case when the $Y_2$
polar-angle is not determined event by event, we integrate the correlations
over the $Y$ polar angle so as to derive the single polar-angle distribution of
one of the $Y$ decay products. The single polar-angle distribution can be
expressed in terms of two polarization estimator functions (PEFs) for
unpolarized $X_2$ particles
\cite{Shelton:2008nq,V.:2016wba,Velusamy:2018ksp,Choi:2018sqc} and eight
polarization estimator functions appearing with non-zero $X_2$ polarization and
accompanied by explicit trigonometric functions of the $Y$ polar angle $\theta$
if the spin values are restricted up to one, i.e. 0, 1/2 and 1. All of the PEFs
are functions in the $X_2$ speed $\beta_2$ in the LAB and the $Y$ speed $\beta$
in the $X_2$RF fixed with the $X_{1,2}$ and $Y$
masses, $m_{1,2}$ and $m$.\\

This paper consists of six main sections and three appendices. After this
introduction part, Section~\ref{sec:wick_helicity_rotation} gives a general
description of the construction of a helicity state of a particle and the
transformation of its related helicity amplitudes by Wick helicity rotation.
Once we derive the $Y$-helicity dependent polar-angle distribution in the
$X_2$RF by integrating the angle-dependent distribution over the $Y$ azimuthal
angle, then we can employ a proper Wick helicity rotation to get the
polar-angle distribution depending directly on the $Y$ helicities in the LAB.
This final angular distribution to be called a Wick helicity rotation
distribution function (WDF) involves only the diagonal elements of the $X_2$
polarization density matrix after an azimuthal-angle integration and this can
be {\it directly} coupled to any polarized decay distribution of the particle
$Y$. In order to facilitate the derivation of WDFs we introduce so-called Wick
helicity rotation spectral functions (WSFs) solely consisting of the pure
kinematic elements for the Wick helicity rotation and the explicitly
angle-dependent part of the helicity amplitude in the $X_2$RF, which plays a
key role in connecting the $X_2$ polarization to the dynamical structure
encoded in the reduced helicity amplitudes and for generating the $Y$
polarization density matrix.
Section~\ref{sec:polar_angle_kinematic_reconstruction} is devoted to combining
the $Y$ density matrix encoding the polarization-dependent angular
distributions of the decay $X_2\to Y X_1$ with the sequential decay $Y\to ab$
into a correlation function of two polar angles, the $Y$ polar-angle $\theta$
in the $X_2$RF and the $a$ polar-angle $\theta_a$, in the $Y$ rest frame. In
Section~\ref{sec:polarization_estimator_functions} we introduce polarization
estimator functions to be used for expressing the single polar-angle
correlation derived by integrating the polar-angle correlation over the $Y$
polar angle so as to cover the situation when the $Y$ polar-angle with respect
to the $X_2$ flight direction cannot be measured experimentally.\\

In Section~\ref{sec:standard_model_beyond_examples} we demonstrate the
formalism for polar-angle correlations explicitly by studying two SM examples,
the sequential process $e^-e^+\to Z\to \tau^-\tau^+ \to
(\rho^-\nu_\tau)\tau^+\to ([\pi^-\pi^0]\nu_\tau)\tau^+$ treating $\tau^+$
inclusively, the sequential process $e^-e^+\to t\bar{t}\to (W^+b)\bar{t}\to
([\ell^+\nu_\ell]b)\bar{t}$, treating $\bar{t}$ inclusively, and one example in
a model beyond the SM with a heavy vector-like top quark, $T\to Zt \to
[\ell^-\ell^+] t$, as one of the characteristic non-standard examples.
Section~\ref{sec:conclusions} contains a summary of our results and concluding
remarks. After that, three appendices collecting a lot of mathematical formulas
to be used in the main text are added.
Appendix~\ref{appendix:wigner_d_functions} is for introducing Wigner
$d$-functions and listing a few properties to be exploited in the present work.
Appendix~\ref{appendix:wick_rotation_distribution_functions} lists all the WDFs
and the $Y$ polarization density matrices in the general form so that they can
be applied to any specific two-stage decays with no further refinements.
Finally,  we present the explicit forms of all the non-trivial polarization
estimator functions and investigate their asymptotic behaviors
in Appendix~\ref{appendix:polarization_estimator_functions}.\\

\section{Wick helicity rotation on helicity states and helicity amplitudes}
\label{sec:wick_helicity_rotation}

A helicity state $|p,j\lambda\rangle$ of a single spin-$j$ particle with
helicity $\lambda$ and its four-momentum $p=(E, \vec{p})=E(1,
\beta\sin\theta\cos\phi, \beta\sin\theta\sin\phi,\beta\cos\theta)$ satisfying
$\beta=|\vec{p}|/E=\sqrt{1-m^2/E^2}$ in a given reference frame is defined by
applying a sequence of boost and rotation transformations to a spin-$j$
angular-momentum eigenstate $|j\lambda\rangle$ with the $z$-axis spin component
$\lambda$ in the rest frame with a fixed coordinate system as
\cite{Leader:2001gr,Wick:1962zz}
\begin{eqnarray}
 |p,j\lambda\rangle
= R_z(\phi)R_y(\theta)L_z(\beta) |j\lambda\rangle\,,
\label{eq:general_helicity_state}
\end{eqnarray}
where the combined operation $R_z(\phi) R_y(\theta)=R(\phi,\theta,0)$ is a
rotation\footnote{It should be noted that this rotation is simpler than the one
introduced in the original paper by Jacob and Wick  corresponding to $R(\phi,
\theta, -\phi)$ in Ref.$\,$\cite{Jacob:1959at}.} taking the $z$-axis into the
direction of $\vec{p}$ with spherical angles $\Omega=(\theta, \phi)$ and
$L_z(\beta)$ is a pure boost along the $z$-axis direction from the rest frame
to the frame where the particle speed is $\beta$. In contrast, the pure Lorentz
transformation by a boost vector $\vec{\beta}$ preserving the assigned
coordinate system is $L(\vec{\beta}) = R(\phi,\theta,0) L_z(\beta)
R^{-1}(\phi,\theta,0)$. For convenience we define the sequence of operations on
the right-hand side in Eq.$\,$(\ref{eq:general_helicity_state}) as an operation
$h(\vec{\beta})$:
\begin{eqnarray}
h(\vec{\beta})= R_z(\phi)R_y(\theta)L_z(\beta) = L(\vec{\beta}) R(\phi,\theta,0)\,,
\end{eqnarray}
with $\vec{\beta}=\vec{p}/E$.
By definition, the helicity quantum number $\lambda$ is the component of the spin
along the momentum $\vec{p}$ and it is a rotationally-invariant quantity.\\

The general theoretical analysis of the polarization of the particle $Y$ of
spin $j$ in the two-body decay $X_2\, \to\, Y\, X_1$ is most transparent in the
$X_2$RF frame if performed in the helicity formalism \cite{Wick:1962zz}. The
decay helicity amplitude can be decomposed in terms of the decay polar and
azimuthal angles for the momentum direction of the particle $Y$ produced in the
$X_2$RF as

\begin{eqnarray}
  D^R_{\sigma_2: \sigma\sigma_1}(\theta, \phi)
= F^{j_2}_{\sigma\sigma_1}\, d^{j_2}_{\sigma_2,\sigma-\sigma_1}(\theta)\,
  e^{i \sigma_2\phi}\quad \mbox{with}\quad |\sigma-\sigma_1|\leq j_2\,,
\label{eq:general_two-body_decay_helicity_amplitude}
\end{eqnarray}
where $j_2$ and $\sigma_2$ are the spin and helicity of the particle $X_2$, and
$\sigma$ and $\sigma_1$ are the helicities of the particles, $Y$ and $X_1$
boson, respectively. For the sake of discussion the $Y$ momentum direction will
be referred to as the production axis in the following. Because of rotational
invariance, the reduced matrix elements $F^{j_2}_{\sigma\sigma_1}$ in
Eq.$\,$(\ref{eq:general_two-body_decay_helicity_amplitude}) containing all the
dynamical information on the decay process is independent of
the $X_2$ helicity $\sigma_2$.\\

The energy $E$ and speed $\beta$ of the particle $Y$ in the decay $X_2\to Y
X_1$ are fixed in the $X_2$RF  with the masses $\{m_2, m, m_1\}$ of the three
particles as
\begin{eqnarray}
E = \frac{m^2_2-m^2_1+m^2}{2m_2} \quad \mbox{and}\quad
\beta = \frac{\lambda^{1/2}(m^2_2, m^2_1, m^2)}{m^2_2-m^2_1+m^2}\,,
\label{eq:Z_energy_momentum}
\end{eqnarray}
with the magnitude of momentum $|\vec{p}|=\beta E$ and the K\"{a}ll\'{e}n
kinematical function $\lambda(x,y,z)=x^2+y^2+z^2-2xy-2yz-2zx=(x-y-z)^2-4yz$
\cite{Kallen:1964lxa}.\\

\begin{figure}[htb]
\centering
\includegraphics[width=11cm, height=7cm]{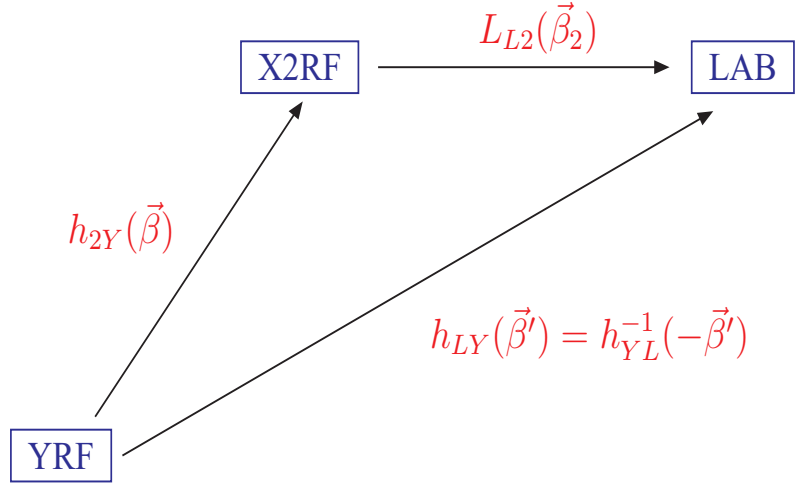}
\caption{{\it A simple diagrammatic description of the Wick helicity rotation describing
              the mismatch between the two helicity coordinate systems of the
              LAB directly reached through one single Lorentz transformation
              $h_{LY}(\vec{\beta}_2)=\Lambda^{-1}_{YL}(-\vec{\beta}_2)$
              and indirectly reached through the Lorentz transformation
              $h_{2Y}(\vec{\beta})$ from the rest frame of $Y$ ($Y$RF) to the $X_2$RF and then
              the pure coordinate-preserving Lorentz transformation
              $L_{L2}(\vec{\beta}_2)$ from the $X_2$RF to the LAB, from the
              $Y$RF. The boost parameter $\beta_2$
              is the $X_2$ speed in the LAB and the parameters, $\beta$ and $\beta'$,
              are the speed of the particle $Y$ in the $X_2$RF and LAB,
              respectively.}}
\label{fig:simple_diagrammatic_description}
\end{figure}

The polarizations of the particle $Y$ determined with respect to the detection
axis of the $Y$ momentum direction in the LAB is related to those in the
$X_2$RF frame by a Wick helicity rotation connecting the two helicity bases
\cite{Wigner_rotation}. The Wick helicity rotation angle $\omega$ is determined
by taking the three sequential operations consisting of the Lorentz
transformation $h_{2Y}(\vec{\beta})$ from the $Y$RF to $X_2$RF, followed by the
pure coordinate-preserving Lorentz transformation $L_{L2}(\vec{\beta}_2)$ from
the $X_2$RF frame to the LAB, and finally the Lorentz transformation
$h_{YL}(-\vec{\beta}')=h^{-1}_{LY}(\vec{\beta}')$ transforming back the system
from the LAB to the $Y$RF as
\begin{eqnarray}
   {\cal R}(\vec{\omega})
=  h^{-1}_{LY}(\vec{\beta}')\, L_{L2}(\vec{\beta}_2)\, h_{2Y} (\vec{\beta})
=  h_{YL}(-\vec{\beta}')\, L_{L2}(\vec{\beta}_2)\, h_{2Y} (\vec{\beta})\,,
\label{eq:wick_helicity_rotation}
\end{eqnarray}
with the $\vec{\omega}$ direction parallel to $\vec{\beta}_2\times
\vec{\beta}$, where $h$, $L$, and ${\cal R}$ are the representation matrices
for the Lorentz transformations. A simple diagrammatic description for the Wick
helicity rotation is shown in
Fig.$\,$\ref{fig:simple_diagrammatic_description}. Explicitly, the $Y$ velocity
$\vec{\beta}'$ in the LAB is related to the $Y$ velocity $\vec{\beta}$ in the
$X_2$RF by
\begin{eqnarray}
  \vec{\beta}'
= \frac{1}{\gamma_2\,(1+\vec{\beta}_2\cdot \vec{\beta})} \,
  \left[\vec{\beta}+\gamma_2 \vec{\beta}_2
       +\frac{\gamma^2_2}{\gamma_2+1} (\vec{\beta}_2\cdot\vec{\beta})\vec{\beta}_2
       \right]
= \frac{\gamma^{-1}_2\vec{\beta}_\bot + \vec{\beta}_\| + \vec{\beta}_2}{
        1+\vec{\beta}_2\cdot \vec{\beta}}\,,
\label{eq:velocity_addition}
\end{eqnarray}
in terms of the velocities, $\vec{\beta}$ and $\vec{\beta}_2$, with
$\gamma_2=1/\sqrt{1-\beta^2_2}$, and the Wick helicity rotation angle $\omega$
defined by the relation in Eq.$\,$(\ref{eq:wick_helicity_rotation}) can be
extracted from the expressions of the standard tangent function
\begin{eqnarray}
  \tan\omega
= \frac{\beta_2 \sqrt{1-\beta^2} \sin\theta}{\beta+\beta_2 \cos\theta}\,,
    \label{eq:wick_rotation_angle}
\end{eqnarray}
and/or those of the sine and cosine functions
\begin{eqnarray}
   \sin\omega
&=& \frac{\beta_2\sqrt{1-\beta^2}\sin\theta}{
    \sqrt{(\beta_2\beta \cos\theta+1)^2-(1-\beta^2_2)(1-\beta^2)}}\,,
   \\
   \cos\omega
&=& \frac{\beta+\beta_2\cos\theta}{
    \sqrt{(\beta_2\beta \cos\theta+1)^2-(1-\beta^2_2)(1-\beta^2)}}\,,
\end{eqnarray}
in terms of the $Y$ speed and polar angle, $\beta$ and $\theta$, in the $X_2$RF
and the $X_2$ speed, $\beta_2$, in the LAB. Similarly, the angle $\omega_1$ of
the Wick helicity rotation for the $X_1$ helicity state and distributions in
the LAB can be obtained from Eq.$\,$(\ref{eq:wick_rotation_angle}) by replacing
$\theta$ by $\pi-\theta$ and $\beta$ by $\beta_1$, the $X_1$ speed in the
$X_2$RF.  \\

There are two extreme kinematic limits for which we do not have to rely on any detailed
information on the boost distributions in practice. Firstly, if the particle $X_2$ is
produced near threshold with $\beta_2\rightarrow 0$, then $\omega \rightarrow 0$
rendering the difference between the production and detection axes negligible.
Secondly, if the mass splitting, $m_2-m_1$, of the particles $X_2$ and $X_1$ is much
larger than $m$, the particle $Y$ is highly boosted with $E$ and $p$ much
larger than $m$ even in the $X_2$ rest frame except for the far backward region with
$\theta$ very close to $\pi$. Naturally, $\omega=0$ if the particle $Y$ is massless,
i.e. $\beta=1$.\\

\begin{figure}[htb]
\centering
\includegraphics[width=13cm, height=8cm]{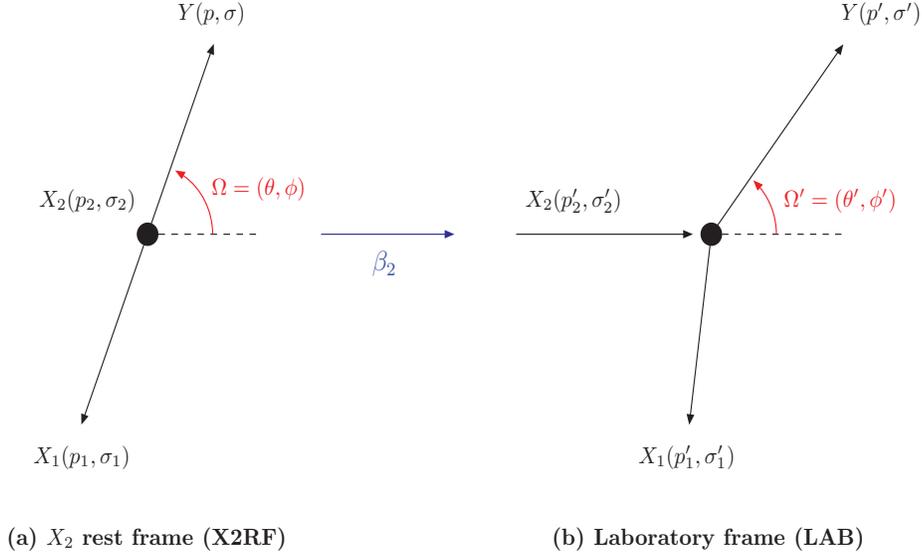}
\caption{{\it Kinematic configurations of the two-body decay $X_2 \to Y X_1$ in (a) the
              $X_2$RF with $p_2=(m_2, \vec{0})$ and (b) in the LAB.
              The unprimed
              $(p_2, q, p_1)$ and $(\sigma_2, \sigma, \sigma_1)$ are the momenta and
              helicities of the $X_2, Y$ and $X_1$ particles in the $X_2$RF while the
              corresponding primed momenta and helicities are defined in the LAB.
              The solid angles, $\Omega=(\theta, \phi)$ and $\Omega' = (\theta', \phi')$,
              which are defined with respect to the $X_2$ momentum direction in the LAB,
              are the polar and azimuthal angles of the particle $Y$ in the $X_2$RF and
              in the LAB, respectively. The boost parameter $\beta_2$, which links the
              $X_2$RF and LAB, is nothing but the $X_2$ speed in the LAB. Note that the
              azimuthal angles of the particle $Y$ and the $X_2$ helicities are
              identical, i.e. $\phi'=\phi$ and $\sigma'_2=\sigma_2$ for the boost.
              }}
\label{fig:kinematic_configurations}
\end{figure}

Let us consider a fixed 3-dimensional spatial coordinate system of the $X_2$RF
with the positive $z$-axis along the $X_2$ momentum direction $\vec{\beta}_2$
in the LAB. In this situation the $X_2$ helicity is invariant under the boost
along the $X_2$ momentum direction from the $X_2$RF to the LAB so that the
helicity states of the particles, $Y$ and $X_1$, in the LAB are given in terms
of the corresponding helicity states in the $X_2$RF by
\begin{eqnarray}
  |p',j\sigma'\rangle
= \sum_{\sigma} d^j_{\sigma',\sigma}(\omega)\, |p, j\sigma\rangle
\quad \mbox{and}\quad
  |p_1',j_1\sigma'_1\rangle
= \sum_{\sigma_1} d^{j_1}_{\sigma'_1,\sigma_1}(\omega_1)\,
  |p_1, j_1\sigma_1\rangle\,.
\label{eq:wick_rotated_helicity_states}
\end{eqnarray}
As a consequence, the decay helicity amplitudes in the LAB for the two-body
decay are related to those in the $X_2$RF through two Wick helicity rotations
on the $Y$ and $X_1$ states as
\begin{eqnarray}
  D^L_{\sigma'_2, \sigma',\sigma'_1} (\theta',\phi')
= \sum_{\sigma\,,\sigma_1}\,
  d^j_{\sigma',\sigma}(\omega)\, d^{j_1}_{\sigma'_1,\sigma_1}(\omega_1)\,
  D^R_{\sigma'_2, \sigma,\sigma_1} (\theta,\phi)\,,
\end{eqnarray}
with $\phi'=\phi$ of the particle $Y$ for this specific Lorentz boost
$L_{L2}(\beta_2 \hat{z})$.
The polar angle $\theta'$ and the energy $E'$ of $Y$ particle in the LAB are
expressed in terms of  to the polar angle $\theta$ as
\begin{eqnarray}
\tan\theta' = \frac{\beta \sin\theta}{\gamma_2(\beta_2 + \beta \cos\theta)}\quad
\mbox{and} \quad
E' = \gamma_2 (1+\beta_2 \beta\, \cos\theta)\, E\,,
\end{eqnarray}
with the explicit forms of $E$ and $\beta$ in
Eq.$\,$(\ref{eq:Z_energy_momentum}). It is noteworthy that, if neither
$\beta_2$ nor $\beta$ is zero, the polar-angle distribution in the $X_2$RF can
be derived directly from the $Y$ energy distribution in the LAB. The kinematic
configurations of the two-body decay $X_2\to Y X_1$ in
the $X_2$RF and LAB are displayed in Fig.$\,$\ref{fig:kinematic_configurations}.\\

In order to describe the impact of the $X_2$ polarization on the $Y$
polarization and angular distribution in the LAB in a general footing, we
introduce the $(2j_2+1)\times (2j_2+1)$ helicity density matrix $\rho^{X_2}$
containing the full information on the $X_2$ polarization and satisfying the
normalization condition ${\rm Tr}(\rho^{X_2})=1$. Integrating over the
azimuthal angle $\phi$ we can obtain the helicity-dependent distribution in the
$X_2$RF as
\begin{eqnarray}
{\cal D}_{\sigma\lambda}(\theta)
= \sum_{\sigma_2} \sum_{\sigma_1}
  \rho^{X_2}_{\sigma_2,\sigma_2}\,
  [d^{j_2}_{\sigma_2, \sigma-\sigma_1}(\theta)\,
  d^{j_2}_{\sigma_2, \lambda-\sigma_1}(\theta)]\,
  F^{j_2}_{\sigma\sigma_1}
  F^{j_2*}_{\lambda\sigma_1}\,.
\label{eq:x2_rest_frame_correlated_distribution}
\end{eqnarray}
By performing the integration\footnote{Even if the azimuthal angle
distributions allow us to make a more detailed spin and angular-correlation
analysis, we focus on the polar-angle distributions, while postponing the full
correlations as our next project.} over the azimuthal angle $\phi'$, which is
identical to $\phi$ under the Lorentz transformation $L_{L2}(\beta_2\hat{z})$,
and taking the sum over the $X_1$ helicity $\sigma'_1$, we can obtain a
fully-correlated and Wick-rotated distribution, from which the $(2j+1)\times
(2j+1)$ polarization density matrix of the particle $Y$ in the LAB can be
derived, as
\begin{eqnarray}
  {\cal D}'_{\sigma'\lambda'}(\omega,\theta)
&=&  \sum_{\sigma,\,\lambda}\,
  [d^j_{\sigma',\sigma}(\omega)\, d^j_{\lambda',\lambda}(\omega)]\,
  {\cal D}_{\sigma\lambda}(\theta)\,, \\
&=&\sum_{\sigma_2} \sum_{\sigma,\,\lambda}   \sum_{\sigma_1}
  \rho^{X_2}_{\sigma_2,\sigma_2}\,
  [d^j_{\sigma',\sigma}(\omega)\, d^j_{\lambda',\lambda}(\omega)]\,
  [d^{j_2}_{\sigma_2, \sigma-\sigma_1}(\theta)\,
  d^{j_2}_{\sigma_2, \lambda-\sigma_1}(\theta)]\,
  F^{j_2}_{\sigma\sigma_1}
  F^{j_2*}_{\lambda\sigma_1}\,,
\label{eq:unnormalized_correlated_distribution}
\end{eqnarray}
to be called {\it Wick helicity rotation distribution functions (WDFs)}
involving only the diagonal components of the density matrix $\rho^{X_2}$,
where the general form in
Eq.$\,$(\ref{eq:general_two-body_decay_helicity_amplitude}) of the two-body
decay helicity amplitude has been taken into account. The polar-angle and
polarization dependent decay width is then given by
\begin{eqnarray}
  \frac{d\Gamma_{\sigma',\lambda'}}{d\cos\theta}
= \gamma_2\beta_2\beta E\, \frac{d\Gamma_{\sigma',\lambda'}}{dE'}
= \frac{\bar{\beta}}{16\pi\gamma_2 m_2}\,
  {\cal D}'_{\sigma'\lambda'} (\omega, \theta)\,,
\end{eqnarray}
where the boost factor $\gamma_2=1/\sqrt{1-\beta^2_2}$ and the abbreviation
$\bar{\beta}= \lambda^{1/2}(1, m^2_1/m^2_2, m^2/m^2_2)$. For the sake of our
discussion, we cast the expression of WDFs in
Eq.$\,$(\ref{eq:unnormalized_correlated_distribution}) into a little shorter
form:
\begin{eqnarray}
  {\cal D}'_{\sigma'\lambda'}(\omega,\theta)
= \sum_{\sigma_2} \sum_{\sigma,\,\lambda}   \sum_{\sigma_1}
  \rho^{X_2}_{\sigma_2,\sigma_2}\,\times\,
  {\cal S}_{\sigma_2;\sigma'\lambda'}^{\sigma\lambda; \sigma_1} (\omega,\theta)
  \times
  F^{j_2}_{\sigma\sigma_1}
  F^{j_2*}_{\lambda\sigma_1}\,,
\label{eq:simplified_correlated_distribution}
\end{eqnarray}
by introducing the following helicity and polar-angle dependent functions to be
called {\it Wick helicity rotation spectral functions (WSFs)}  as
\begin{eqnarray}
  {\cal S}_{\sigma_2;\sigma'\lambda'}^{\sigma\lambda; \sigma_1} (\omega,\theta)
= [d^j_{\sigma',\sigma}(\omega)\, d^j_{\lambda',\lambda}(\omega)]\,\times\,
  [d^{j_2}_{\sigma_2, \sigma-\sigma_1}(\theta)\,
  d^{j_2}_{\sigma_2, \lambda-\sigma_1}(\theta)]
\label{eq:wick_distribution_functions}
\end{eqnarray}
The averages of WSFs over the polar-angle $\theta$ to be named {\it Wick
helicity rotation spectral elements (WSEs)} is given by
\begin{eqnarray}
  \langle {\cal S}_{\sigma_2;\sigma'\lambda'}^{\sigma\lambda; \sigma_1} \rangle
= \frac{1}{2} \int^{1}_{-1}\,
 {\cal S}^{\sigma\lambda; \sigma_1}_{\sigma_2;\sigma'\lambda'} (\omega,\theta)\,\,
 d\cos\theta\,.
\label{eq:wick_spectral_functions}
\end{eqnarray}
The WSFs satisfy the normalization conditions
\begin{eqnarray}
  {\rm Tr}(S^{\sigma\lambda; \sigma_1}_{\sigma_2})
\equiv \sum_{\sigma'} S^{\sigma\lambda; \sigma_1}_{\sigma_2; \sigma'\sigma'}
= \delta_{\sigma\lambda}\,[d^{j_2}_{\sigma_2,\sigma-\sigma_1}(\theta)]^2\,,
\end{eqnarray}
with no Wick helicity rotation effects, leading to a simple normalization for
the WSEs as
\begin{eqnarray}
{\rm Tr}(\langle {\cal S}^{\sigma\lambda; \sigma_1}_{\sigma_2} \rangle )
\equiv \sum_{\sigma'} \langle
    S^{\sigma\lambda; \sigma_1}_{\sigma_2; \sigma'\sigma'}\rangle
= \left\{ \begin{array}{cl}
          \delta_{\sigma\lambda}/(2j_2+1) & \mbox{if}\ \ |\sigma-\sigma_1|\leq j_2\,,
           \\[2mm]
          0 & \mbox{if}\ \ |\sigma-\sigma_1| > j_2\,,
          \end{array}
          \right.
\end{eqnarray}
that is independent of the $X_2$ helicity $\sigma_2$. \\

The normalized polar-angle dependent distribution $W'$ and the integrated
polarization density matrix $\rho^Y$ of the particle $Y$ are obtained from the
WDFs in Eq.$\,$(\ref{eq:unnormalized_correlated_distribution}) as
\begin{eqnarray}
  W'_{\sigma'\lambda'}(\theta)
= \frac{{\cal D}'_{\sigma'\lambda'}(\omega,\theta)}{
   {\rm Tr}[\langle {\cal D}'\rangle]}
\quad\ \  \mbox{and} \quad\ \
  \rho^Y_{\sigma'\lambda'}
= \frac{\langle {\cal D}'_{\sigma'\lambda'}(\omega,\theta)\rangle}{
        {\rm Tr}[\langle {\cal D}'\rangle]}\,,
\label{eq:y_density_matrices_lab}
\end{eqnarray}
satisfying the normalization conditions, ${\rm Tr}[\langle W'\rangle] =1$ and
${\rm Tr}(\rho^Y)=1$. They will be combined later with the polarized $Y$ decay
distributions, for correlated polar-angle distributions and single polar-angle
distributions.  The explicit expressions of the matrix elements will be
presented in detail for a specific set of decay processes later in
Section~$\,$\ref{sec:standard_model_beyond_examples} and the polarization
density matrices  $\rho^Y$ are listed in their general form for the cases with
particle spins up to one in
Appendix~\ref{appendix:wick_rotation_distribution_functions}.\\

In passing we note that the partial decay width $\Gamma[X_2\to Y X_1]_{LAB}$ in
the LAB is obtained by summing up the diagonal elements of the distribution
matrix in Eq.$\,$(\ref{eq:unnormalized_correlated_distribution}) over the $Y$
helicity $\sigma'$ and integrating it over the polar angle $\theta$ as
\begin{eqnarray}
  \Gamma[X_2\to Y X_1]_{LAB} = \frac{1}{\gamma_2} \Gamma[X_2\to Y X_1]
= \frac{\bar{\beta}}{8\pi \gamma_2 m_2} \frac{1}{(2j_2+1)}
  {\sum_{\sigma,\sigma_1}}' |F^{j_1}_{\sigma\sigma_1}|^2\,,
\end{eqnarray}
with the boost factor $\gamma_2$ of the particle $X_2$ of speed $\beta_2$ in
the LAB, reflecting time dilation. The prime on the summation notation
indicates that the sum
is taken only when $|\sigma-\sigma_1|\leq j_2$ is satisfied.\\

\section{Polar-angle correlations and their reconstruction}
\label{sec:polar_angle_kinematic_reconstruction}

Extracting efficiently the essential information on the $X_2$ polarization and
the dynamics of the two-body decay $X_2\to YX_1$ encoded in the $Y$ density
matrix $\rho^Y(\theta)$ in Eq.$\,$(\ref{eq:y_density_matrices_lab})
require exploiting $Y$-polarization sensitive decays. \\

With no serious loss of generality, we assume that $Y$ decays into two
particles, a particle $a$ of mass $m_a$ and spin $j_a$ and a particle $b$ of
mass $m_b$ and spin $j_b$ of which the helicity amplitude can be written in the
$Y$ rest frame as
\begin{eqnarray}
  D_{\sigma;\sigma_a\sigma_b}(\theta_a,\phi_a)
= H^j_{\sigma_a\sigma_b} d^j_{\sigma,\sigma_a-\sigma_b}(\theta_a)\,
  e^{i\sigma\phi_a}
\,,
\end{eqnarray}
where the polar and azimuthal angles $\theta_a$ and $\phi_a$ define the
momentum direction of the particle $a$ in a coordinate system with the positive
$z$ axis
along the $Y$ flight direction in the LAB.\\

After combining the $Y$ production and decay amplitudes and integrating the
combined distribution over the azimuthal angle $\phi_a$, we obtain a correlated
distribution of two polar angles, $\theta$ and $\theta_a$, as
\begin{eqnarray}
  \frac{d^2{\cal C}}{d\cos\theta\,d\cos\theta_a}
= \frac{(2j+1)}{4}\sum^j_{\sigma'=-j} W'_{\sigma'\sigma'}(\theta)\,\,
                 \rho^D_{\sigma'\sigma'}(\theta_a)\,,
\label{eq:two_dimension_correlation_distribution}
\end{eqnarray}
with the $Y$ decay density matrix $\rho^Y$ defined in terms of the decay
helicity amplitudes as
\begin{eqnarray}
   \rho^D_{\sigma'\sigma} (\theta_a)
 = \frac{\sum_{\sigma_a\sigma_b} [d^j_{\sigma', \sigma_a-\sigma_b}(\theta_a)]^2
|H^j_{\sigma_a\sigma_b}|^2}{\sum_{\sigma_a\sigma_b} |H^j_{\sigma_a\sigma_b}|^2}
\,,
\label{eq:y_decay_density_matrix}
\end{eqnarray}
The polar-angle correlation in
Eq.$\,$(\ref{eq:two_dimension_correlation_distribution}) encodes the full
information on the dynamics of the two-stage decay $X_2\to Y X_1\to (ab) X_1$
that can be extracted through measuring the polar angles, $\theta$ and
$\theta_a$, experimentally. Here, we emphasize that the correlation function
depends also on the $X_2$ speed $\beta_2$ as well. In the following analysis,
we assume that the speed $\beta_2$ is determined event by event or the
$\beta_2$-dependent distribution is known already so that it can be folded with
the correlation function for a full-fledged
distribution.\\

\begin{figure}[htb]
\centering
\includegraphics[width=10cm, height=8cm]{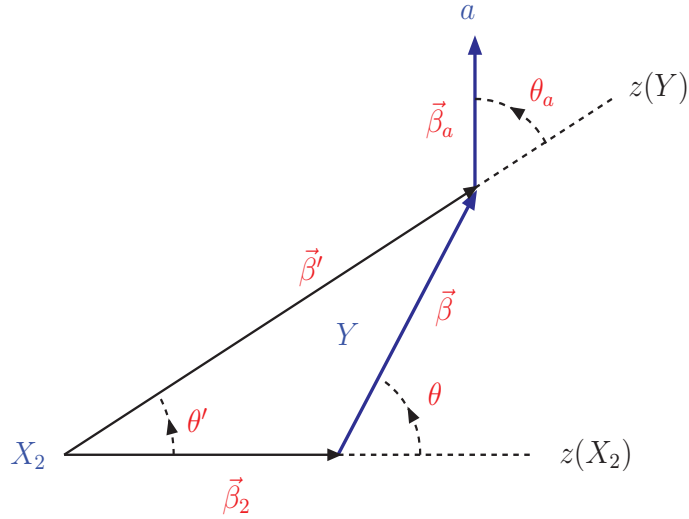}
\caption{{\it A simple diagram describing the relation among the polar angles,
              $\theta$ in the $X_2$RF and $\theta'$ in the LAB, and the polar angle
              $\theta_a$ of the particle in the $Y$ rest frame from the decay
              $Y\to a+b$. $\vec{\beta}_2$ is the $X_2$ velocity in the LAB and
              $\vec{\beta}(\vec{\beta}')$ and $\theta(\theta')$ are the $Y$ velocity
              and polar angle in the $X_2$RF (LAB). Also, $\vec{\beta}_a$ and $\theta_a$
              are the $a$ velocity and polar angle in the $Y$RF. $z(X_2)$ and $z(Y)$
              denote the $X_2$ and $Y$ flight directions in the LAB.
              Note that the relative orientation between the plane formed by
              $\vec{\beta}_2$ and $\vec{\beta}$ and the plane formed by $\vec{\beta}'$
              and $\vec{\beta}_1$ can be arbitrary. In this light, all the azimuthal
              angles are not displayed for simplicity, because they are irrelevant
              to the polar-angle correlations under our considerations.}}
\label{fig:kinematics_two_stage_decay}
\end{figure}

The most straightforward way of determining the polar angles $\theta$ and
$\theta_a$ is through the measurement of the $Y$ and $a$ energies, $E'$ and
$E'_a$, in the LAB, as they satisfy the relations:
\begin{eqnarray}
    \cos\theta
&=& \frac{1}{\gamma_2\beta_2\gamma\beta}
    \left[\frac{E'}{m}-\gamma_2\gamma\right]\,,
    \label{eq:costheta_E_prime} \\
    \cos\theta_a
&=& \frac{1}{\gamma'\beta'\gamma_a\beta_a}
    \left[\frac{E'_a}{m_a}-\gamma'\gamma_a\right]\,,
    \label{eq:costheta_a_E_a_prime}
\end{eqnarray}
where the $Y$ boost factors, $\gamma'$ and $\beta'$, in the LAB are given in
terms of the $X_2$ boost factors in the LAB, $\beta_2$ and $\gamma_2$, and the
$Y$ polar angle $\theta$ and $Y$ boost factors $\gamma$ and $\beta$ in the
$X_2$RF by
\begin{eqnarray}
\gamma' = \frac{E'}{m}=\gamma_2\gamma\, (1+\beta_2\beta\cos\theta)
\quad\mbox{and}\quad
\beta'=\sqrt{1-1/\gamma'^2}\,,
\end{eqnarray}
as can be checked with  Eq.$\,$(\ref{eq:costheta_E_prime}) The value of
$\gamma'$ varies between $\gamma_2\gamma (1-\beta_2\beta\cos\theta)$ and
$\gamma_2\gamma (1+\beta_2\beta\cos\theta)$ and, for a given value of
$\gamma'$, the allowed range of the $a$ boost factor $\gamma'_a= E'_a/m_a$ is
between $\gamma'_{a{\rm min}} = \gamma'\gamma_a (1-\beta' \beta_a)$ and
$\gamma'_{a{\rm max}} = \gamma'\gamma_a (1+ \beta' \beta_a)$ with $\gamma_a =
(m^2+m^2_a-m^2_b)/2 m m_a$. A simple diagrammatic description of the kinematic
relations among angles and boost parameters is shown in
Fig.$\,$\ref{fig:kinematics_two_stage_decay}. Note that $\vec{\beta}'$ is not a
simple vector sum of $\vec{\beta}_2$ and $\vec{\beta}$ but a complicated
combination of them as shown in
Eq.$\,$(\ref{eq:velocity_addition}).\\

If the polar angle $\theta$ in the $X_2$RF cannot be measured but the
four-momentum of $Y$ or one of the $Y$ decay products can be fully
reconstructed, then we can still use the one-dimensional $\theta_a$
distribution derived by integrating the 2-dimensional distribution in
Eq.$\,$(\ref{eq:two_dimension_correlation_distribution}) over the polar angle
$\theta$. In general there are thirteen independent functions of the Wick angle
$\omega$ and $\theta$ to be integrated over the angle $\theta$. Three of them
are rather trivial as they depend simply on $\theta$. The remaining ten
integrated functions, which we call polarization estimator functions (PEFs),
will be classified and described in detail in the next section,
while their expressions with spins up to one are listed in
Appendix~\ref{appendix:polarization_estimator_functions}.  \\

\section{Polarization estimator functions}
\label{sec:polarization_estimator_functions}

If the spin of the particle $Y$ is $j=0$, there are no Wick helicity rotation
effects and so the production-decay correlation distribution of the particle
$X_2$ is simply given by the $\theta$-dependent function
\begin{eqnarray}
  {\cal D}'(\omega,\theta) = {\cal D}'(0,\theta)
= \sum_{\sigma_2} {\sum_{\sigma_1}}' \rho^{X_2}_{\sigma_2, \sigma_2}
  \left[d^{j_2}_{\sigma_2,-\sigma_1}(\theta)\right]^2
  |F^{j_2}_{\sigma_1}|^2\,,
\end{eqnarray}
with the restriction $|\sigma'_1|\leq j_2$.
Furthermore, for a spin-0 particle $X_2$ the helicity density matrix $\rho^{X_2}$
is trivially one, resulting in no production-decay correlations at all.\\

On the contrary, non-trivial Wick helicity rotation effects are developed for
non-zero $Y$ spins. Although the formalism given in
Section~\ref{sec:wick_helicity_rotation} can be applied to any spin
combination, we consider the cases with spin values, $0, 1/2$ and $1$ for
showing the {\it diagonal} correlated distributions ($\sigma'=\lambda'$) - WDFs
and WSFs explicitly in the present work as they are directly related with the
sequential polar-angle decay distributions of the particle $Y$ after
azimuthal-angle
integration.\\

Applying the coupling rule of Wigner $d$-functions in
Eq.$\,$(\ref{eq:d_coupling_rule_2}) we can rewrite the diagonal WDFs ${\cal
D}_{\sigma'\sigma'}$ in the form as
\begin{eqnarray}
  {\cal D}'_{\sigma'\sigma'}(\omega,\theta)
=  \sum_{J=0}^{2j} \sum_{J_2=0}^{2j_2}
   \sum_{\sigma,\lambda=-j}^j\,\mathcal{C}^{JJ_2}_{\sigma\lambda;\sigma'}\,\,
  [d^{J}_{0,\,\sigma-\lambda}(\omega)\, d^{J_2}_{0,\,\sigma-\lambda}(\theta)]\,,
\label{eq:coupled_WDFs}
\end{eqnarray}
where the coefficients $\mathcal{C}^{JJ_2}_{\sigma\lambda;\sigma'}$ are
determined by a combination of the elements of the density matrix $\rho^{X_2}$,
the reduced decay helicity amplitudes $F^{j_2}_{\sigma\sigma_1}$ and
$F^{j_2*}_{\lambda\sigma_1}$ and four Clebsch-Gordan coefficients. The general
form of $d^J_{0\lambda}(\beta)$ expressed in terms of the standard Legendre
polynomial $P_J(\cos\beta)$ by the formula \cite{Rose:2011aa}
\begin{eqnarray}
  d^J_{0\lambda}(\beta)
= 2^\lambda \left(\frac{\cos\beta}{2}\right)^\lambda
            \left(\frac{\sin\beta}{2}\right)^\lambda
            \sqrt{\frac{(J-\lambda)!}{(J+\lambda)!}}
            \frac{d^\lambda}{d\cos\beta^\lambda} P_J(\cos\beta)\,,
\end{eqnarray}
for integral $J$ and $\lambda$.
Taking into account the expressions of six $d^J_{0\lambda}$ functions up to $J=2$
with $\lambda\geq 0$:
\begin{eqnarray}
&& d^0_{00}(\beta) = 1; \ \ d^1_{00}(\beta) = \cos\beta, \ \
   d^1_{01}(\beta) = \frac{\sin\beta}{\sqrt{2}}; \\
&& d^2_{00}(\beta) = \frac{1}{2}(3\cos^2\beta-1),\ \
   d^2_{01}(\beta) = \sqrt{\frac{3}{2}} \cos\beta\sin\beta,\ \
   d^2_{02}(\beta) = \sqrt{\frac{3}{8}} (1-\cos^2\beta)\,,
\end{eqnarray}
and three additional $d$-functions with negative $\lambda$ values derived with
the relation $d^J_{0\lambda}(\beta)=(-1)^\lambda d^J_{0,-\lambda}$, ten
non-trivial $\beta_2$ and $\beta$ dependent polarization estimator functions
(PEFs) can be formed:
\begin{eqnarray}
\begin{array}{lll}
  \langle \cos\omega\rangle\,, \quad\quad\quad
& \langle \cos\omega\cos\theta\rangle\,, \quad\quad\quad
& \langle \cos\omega\cos 2\theta\rangle\,, \quad\quad \\[1mm]
  \langle \cos^2\omega\rangle\,,
& \langle \cos^2\omega\cos\theta\rangle\,,
& \langle \cos^2\omega\cos^2\theta\rangle\,, \\[1mm]
& \langle \sin\omega\sin\theta\rangle\,,
& \langle \sin\omega\sin2\theta\rangle\,,   \\[1mm]
& \langle \cos\omega\sin\omega\sin\theta\rangle\,,
& \langle \cos\omega\sin\omega\cos\theta\sin\theta\rangle\,,  \\
\end{array}
\end{eqnarray}
where the bracket stands for the average over the polar-angle $\theta$ defined
as
\begin{eqnarray}
  \langle F(\omega,\theta) \rangle
= \frac{1}{2}\int^1_{-1}\, F(\omega,\theta) \, d \cos\theta\,,
\end{eqnarray}
for any function $F(\omega,\theta)$ dependent on $\theta$ implicitly as well as
explicitly.\footnote{As can be checked with Eq.$\,$(\ref{eq:coupled_WDFs}),
every $\sin\omega$ due to the interference of two $Y$ states of different
helicities in the $X_2$RF is always accompanied by $\sin\theta$ so that
$\langle \sin\omega\rangle$ and $\langle\cos\omega\sin\omega\rangle$ cannot
show up.} Two PEFs, $\langle \cos\omega\rangle$ and $\langle
\cos^2\omega\rangle$ with no explicit $\theta$-dependence, were already
introduced in
Refs.~\cite{Shelton:2008nq,V.:2016wba,Velusamy:2018ksp,Choi:2018sqc}, which
appear even in the case of unpolarized $X_2$. The detailed expressions and the
properties of all the ten non-trivial polarization estimator functions are
listed and described in detail
in Appendix~\ref{appendix:polarization_estimator_functions}.\\

\begin{table}[ht]
\centering
\begin{tabular}{|c|c|c|c|c|c|c|c|c|c|}
\multicolumn{10}{c}{ }  \\[-2mm]
\multicolumn{10}{c}{\color{black} \large Polarization Estimator Functions}\\[2mm]
\hline
{ } & \multicolumn{3}{c|}{ }
    & \multicolumn{3}{c|}{ }
    & \multicolumn{3}{c|}{ }\\[-2mm]
$X_s$ spin & \multicolumn{3}{c|}{$j_2=0$}
           & \multicolumn{3}{c|}{$j_2=\frac{1}{2}$}
           & \multicolumn{3}{c|}{$j_2=1$}\\[2mm]
\hline\hline
 &  &  &  &  &  &  &  &  &  \\[-2mm]
$Y\& X_1$ spins: $\|j j_1\|$
               & $\|\frac{1}{2}\frac{1}{2}\|$  & $\|\langle 10\|$   & $\|11\|$
               & $\|\frac{1}{2}0\|$  & $\|\frac{1}{2} 1\|$
               & $\|1\frac{1}{2}\|$
               & $\|\frac{1}{2}\frac{1}{2}\|$ & $\|10\|$ & $\|11\|$
               \\[2mm] \hline\hline
 &  &  &  &  &  &  &  &  &  \\[-3mm]
$\langle \cos\omega\rangle$   & $\surd$ &         & $\surd$
                              & $\surd$ & $\surd$ & $\surd$
                              & $\surd$ & $\surd$ & $\surd$
\\[2mm] \hline
 &  &  &  &  &  &  &  &  &  \\[-3mm]
$\langle \cos^2\omega\rangle$ &         & $\bullet$ & $\bullet$
                              &         &           & $\bullet$
                              &         & $\bullet$ & $\bullet$
\\[2mm] \hline
 &  &  &  &  &  &  &  &  &  \\[-3mm]
$\langle \cos\omega\cos\theta\rangle$ &            &           &
                                      & $\bullet$  & $\bullet$ & $\bullet$
                                      & $\bullet$  & $\bullet$ & $\bullet$
\\[2mm] \hline
 &  &  &  &  &  &  &  &  &  \\[-3mm]
$\langle \cos\omega\cos^2\theta\rangle$ &          &         &
                                        &          &         &
                                        & $\surd$  & $\surd$ & $\surd$
\\[2mm] \hline
 &  &  &  &  &  &  &  &  &  \\[-3mm]
$\langle \cos^2\omega\cos\theta\rangle$
                                        &          &         &
                                        &          &         & $\surd$
                                        &          & $\surd$ & $\surd$
\\[2mm] \hline
 &  &  &  &  &  &  &  &  &  \\[-3mm]
$\langle \cos^2\omega\cos^2\theta\rangle$
                                        &          &           &
                                        &          &           &
                                        &          & $\bullet$ & $\bullet$
\\[2mm] \hline
 &  &  &  &  &  &  &  &  &  \\[-3mm]
$\langle \sin\omega\sin\theta\rangle$ &            &           &
                                      & $\bullet$  & $\bullet$ & $\bullet$
                                      & $\bullet$  & $\bullet$ & $\bullet$
\\[2mm] \hline
 &  &  &  &  &  &  &  &  &  \\[-3mm]
$\langle \sin\omega\cos\theta\sin\theta\rangle$
                                        &          &         &
                                        &          &         &
                                        & $\surd$  & $\surd$ & $\surd$
\\[2mm] \hline
 &  &  &  &  &  &  &  &  &  \\[-3mm]
$\langle \cos\omega\sin\omega\sin\theta\rangle$
                                        &          &         &
                                        &          &         & $\surd$
                                        &          & $\surd$ & $\surd$
\\[2mm] \hline
 &  &  &  &  &  &  &  &  &  \\[-3mm]
$\langle \cos\omega\sin\omega\cos\theta\sin\theta\rangle$
                                        &          &           &
                                        &          &           &
                                        &          & $\bullet$ & $\bullet$
\\[2mm] \hline
\end{tabular}
\vskip 0.5cm
\caption{\it List of non-trivial polarization estimator functions which may contribute
          to the polar-angle
          averages of WDFs in the decay process $j_2\to j + j_1$ involving all
          the spin values up to 1. The notation $\|jj_1\|$ stands for the final
          state with $Y$ of spin $j$ and $X_1$ of spin $j_1$ and the symbol
          $\bullet$ indicates the polarization estimator functions may shows up
          in the corresponding decay mode, but the symbol $\surd$ implies that the
          corresponding PEF can appear only when parity is violated in
          the two-body decay. Note that every term involving an odd number of
          cosines can show up with parity-violating interactions.}
\label{tab:polarization_estimator_functions}
\end{table}

Table~\ref{tab:polarization_estimator_functions} shows all the non-trivial PEFs
that may contribute to the polar-angle averages of WDFs in the decay process
with the spin combinations of $j_2\to j + j_1$ involving all the spin values up
to one. The notation $\|jj_1\|$ stands for the final state with $Y$ of spin $j$
and $X_1$ of spin $j_1$. The symbol $\bullet$ indicates the polarization
estimator functions may shows up in the corresponding decay mode, but the
symbol $\surd$ implies that the corresponding PEF can appear only when
parity is violated in the two-body decay.\\

\section{Examples of the Wick helicity rotation in the Standard Model and beyond}
\label{sec:standard_model_beyond_examples}

Spin has played a dramatic role in the field of elementary particle physics,
acting as a powerful tool in the confirmation and verification of particle
physics theories, especially in numerous tests of the SM since its birth about
fifty years ago
\cite{Glashow:1961tr,Weinberg:1967tq,Salam:1968rm,Fritzsch:1973pi}. In this
section, we apply the formalism developed in the previous sections to two
well-known SM cases and one non-standard case with a new heavy vectorlike top
quark, eventually deriving the two-stage polar-angle correlations in their full
form. On the other hand, we present a few simple numerical analyses while
postponing more comprehensive numerical studies as a next project.\\

\subsection{The process \boldmath{$e^-e^+\to Z\to \tau^-\tau^+$} followed by
            \boldmath{$\tau^-\to \rho^-\nu_\tau$} and
            \boldmath{$\rho^-\to \pi^-\pi^0$}}
\label{subsec:tau_pair_production}

As a characteristic example of the key decay mode with the spin combination of
$1/2\to 1+ 1/2$, we consider the following three-stage sequential processes of
the SM, established with exquisite precision experimentally at SLAC and LEP and
in the SM \cite{ALEPH:2005ab}:
\begin{eqnarray}
e^-e^+ \to Z \to \tau^-\tau^+ \to (\rho^-\nu_\tau)\tau^+
             \to ([\pi^-\pi^0]\nu_\tau)\tau^+\,,
\label{eq:ee_Z_tautau_rhonutau_pipinutau}
\end{eqnarray}
where $\tau^+$ is assumed to be inclusively measured. The key chain for our
analysis in Eq.$\,$(\ref{eq:ee_Z_tautau_rhonutau_pipinutau}) is the two-body
decay $\tau^-\to\rho^-\nu_\tau$, one of the main $\tau$ decay modes
\cite{Kuhn:1982di,Alemany:1991ki,Kuhn:1991cc,Kuhn:1992nz,Davier:1992nw,Hagiwara:1989fn
}. The $\tau$-pair production process proceeds at the tree level through two
$s$-channel $\gamma$ and $Z$ exchanges. On the $Z$-boson pole, the contribution
from $\gamma$ exchange is of order $\Gamma_Z/m_Z\sim 0.027$ compared with that
of $Z$ exchange \cite{Tanabashi:2018oca} so that the $\gamma$-exchange
contribution can be neglected with good approximation, although it can be
included easily if necessary. The sequential process in
Eq.$\,$(\ref{eq:ee_Z_tautau_rhonutau_pipinutau})
is then viewed as a typical physical process of resonance formation and decay. \\

The Feynman rules of the $eeZ$ and $Z\tau\tau$ vertices consist of vector and
axial-vector structures:
\begin{eqnarray}
    \langle Z | e^-e^+\rangle^\mu
&=& -ig_Z\,\gamma^\mu (v_e-a_e\gamma_5)\,, \\
    \langle \tau^-\tau^+| Z_\mu \rangle_\mu
&=& -ig_Z \gamma_\mu\, (v_\tau-a_\tau\gamma_5)\,,
\end{eqnarray}
with $g_Z=e/c_W s_W$ and the abbreviations, $s_W=\sin\theta_W$ and $c_W =
\cos\theta_W$, of the weak-mixing angle $\theta_W$. In the SM,
$v_e=v_\tau=s^2_W-1/4$ and $a_e=a_\tau= -1/4$. Apart from a function related to
the energy-dependent $Z$ propagator and an azimuthal-angle dependent phase as
well as a common gauge coupling $g_Z$, the helicity amplitude of the
$\tau^-\tau^+$ pair production in $e^-e^+$ collisions can be written in the
form
\begin{eqnarray}
  T_{\sigma_e\bar{\sigma}_e;\sigma_\tau\bar{\sigma}_\tau}(\Theta)
=   P_{\sigma_e\bar{\sigma}_e}\,
    D_{\sigma_\tau\bar{\sigma}_\tau}\,\,
    d^1_{\sigma_e-\bar{\sigma}_e, \sigma_\tau-\bar{\sigma}_\tau}(\Theta)\,.
\end{eqnarray}
The labels, $(\sigma_e,\bar{\sigma}_e)$ and $(\sigma_\tau, \bar{\sigma}_\tau)$,
refer to the helicities of the relevant particles $e^\mp$ and $\tau^\mp$, and
$P_{\sigma_2\bar{\sigma}_e}$ and $D_{\sigma_\tau\bar{\sigma}_\tau}$ measure the
helicity amplitudes for $e^-e^+\to Z$ and $Z\to\tau^-\tau^-$, respectively.\\

If the electron mass $m_e$ is neglected, then the electron and positron must
have opposite helicity, yielding two surviving $e^-e^+\to Z$ helicity
amplitudes as
\begin{eqnarray}
P_{\pm\mp} = v_e \mp a_e\ \ \mbox{and}\ \ P_{\pm\pm}=0\,.
\end{eqnarray}
The latter vanishing result is due to chirality preservation in the limit of
$m_e=0$. On the other hand, with non-zero $\tau$ mass $m_\tau$, the decay part
consists of four helicity amplitudes
\begin{eqnarray}
D_{\pm\mp} &=& v_\tau\mp \beta_\tau a_\tau\,, \\
D_{\pm\pm} &=& \frac{1}{\sqrt{2}} \sqrt{1-\beta^2_\tau}\, v_\tau
                   =  \sqrt{2}\,\frac{m_\tau}{m_Z}\, v_\tau\,,
\end{eqnarray}
with the $\tau^-$ speed $\beta_\tau = \sqrt{1-4m^2_\tau/m^2_Z}$ in the $e^-e^+$
CM frame. Note that $\beta_\tau\simeq 1$ up to per-mille precision with
$m_Z=91\,{\rm GeV}$ and $m_\tau=1.78\,{\rm GeV}$ on the $Z$-boson
pole, i.e. the produced $\tau^\mp$ is highly relativistic. \\

For the sake of notation, we introduce two asymmetry parameters:
\begin{eqnarray}
A_e = \frac{2 v_e a_e}{v^2_e+a^2_e} \quad \mbox{and}\quad
A_\tau = \frac{2v_\tau a_\tau}{v^2_\tau + a^2_\tau}\,,
\end{eqnarray}
and two polarization-dependent quantities
\begin{eqnarray}
&& \xi_1 = -P_{e^-} P_{e^+} - A_e (P_{e^-}-P_{e^+})\,,\\[1mm]
&& \xi_2 = -(P_{e^-}-P_{e^+}) - A_e P_{e^-}P_{e^+}\,.
\end{eqnarray}
Folding the $e^-$ and $e^+$ diagonal $2\times 2$ polarization density matrices
$\rho_{e^\pm} = \frac{1}{2}\, {\rm diag}(1+P_{e^\pm}, 1-P_{e^\pm})$ in the
$(1/2, -1/2)$ helicity basis for longitudinally polarized electron and positron
beams, with the squares of the transition amplitudes and then summing them over
the $\tau^+$ helicities we have the differential cross section given by
\begin{eqnarray}
 \frac{1}{\sigma_{unp}} \frac{d\sigma_{pol}}{d\cos\Theta}
= \frac{3}{2}\frac{(1+\xi_1)[1+\cos^2\Theta+(1-\beta^2_\tau) (\eta_\tau-\cos^2\Theta)]
       +2(A_e+\xi_2)\, A_\tau \beta_\tau \cos\Theta}{
       3+\beta^2_\tau+3\eta_\tau(1-\beta^2_\tau)}\,,
\end{eqnarray}
with a ratio $\eta_\tau=(v^2_\tau-a^2_\tau)/(v^2_\tau+a^2_\tau)$ of the
vector and axial-vector couplings in addition to $A_e$ and $A_\tau$.\\

The angular-dependent $\tau^-$ polarization $P_\tau(\Theta)$ in the $e^-e^+$ CM
frame of a $\tau^-$ moving with speed $\beta_\tau$ and polar angle $\Theta$
reads
\begin{eqnarray}
  P_\tau (\Theta)
&=&  -\frac{(1+\xi_1)A_\tau \beta_\tau (1+\cos^2\Theta)
         +(A_e +\xi_2)[1+\eta_\tau +(1-\eta_\tau)\beta^2_\tau] \cos\Theta}{
        (1+\xi_1)[1+\cos^2\Theta+(1-\beta^2_\tau) (\eta_\tau-\cos^2\Theta)]
        + 2(A_e+\xi_2)A_\tau \beta_\tau \cos\Theta}\,, \\
&& \Rightarrow \ \
    -\frac{(1+\xi_1)A_\tau (1+\cos^2\Theta)
         +2(A_e +\xi_2)\cos\Theta}{
        (1+\xi_1)(1+\cos^2\Theta)
         + 2(A_e+\xi_2) A_\tau\cos\Theta}
        \quad\ \ \mbox{as}\quad \beta_\tau \to 1\,.
\end{eqnarray}
From the statistics point of view it is worthwhile to deal with the degree of
polarization multiplied with the angular distribution and then integrated over
the polar angle
\begin{eqnarray}
  P_\tau
= \frac{1}{\sigma_{pol}} \int\, P_\tau (\theta)
  \frac{d\sigma_{pol}}{d\cos\Theta}\, d\cos\Theta
= -\frac{4\beta_\tau}{3+\beta^2_\tau + 3\eta_\tau(1-\beta^2_\tau)}\, A_\tau
  \ \ \Rightarrow \ \ -A_\tau \ \ \mbox{as}\ \ \beta_\tau\to 1\,,
\end{eqnarray}
that turns out to be independent of the initial electron and positron
polarization as well as
the $ee Z$ couplings.\\

Let us now discuss how the $\pi^-$ spectra arising from the two-stage two-body
decays
\begin{eqnarray}
\tau^- \to \rho^-\nu_\tau \to (\pi^-\pi^0)\nu_\tau
\end{eqnarray}
can be used to determine the polarization of the vector meson $\rho^-$, acting
as a polarization analyzer of the parent particle $\tau^-$
\cite{Hagiwara:1989fn}. The decay mode $\tau^-\to\rho^-\nu_\tau$ accounts for
approximately 22\% of all $\tau$ decays. Adopting the helicity formalism, the
transition amplitude of the process $\tau^-\to\rho^- \nu_\tau$ are given by
\begin{eqnarray}
  A[\tau^-(p,\sigma_\tau)\to\rho^-(q,\lambda) \nu_\tau(k,-)]
= \sqrt{2} G_F g_\rho\,
  [\bar{u}(k,-) \gamma^\mu P_L u(p,\sigma_\tau)]\,
  \epsilon^*_\mu(q,\lambda)\,,
\end{eqnarray}
with $P_L=(1-\gamma_5)/2$. The helicity amplitude can be cast into the
normalized form in the $\tau^-$ rest frame as
\begin{eqnarray}
    D_{\sigma_\tau;-}(\theta_\rho,\phi_\rho)
&=& \frac{\sqrt{2}m_\rho}{\sqrt{m^2_\tau+2m^2_\rho}}\,
    d^{1/2}_{\sigma_\tau, -1/2}(\theta_\rho)\, e^{i\sigma_\tau\phi_\rho}\,,
    \\
    D_{\sigma_\tau;0}(\theta_\rho,\phi_\rho)
&=& \frac{m_\tau}{\sqrt{m^2_\tau+2m^2_\rho}}\,\,
    d^{1/2}_{\sigma_\tau, 1/2}(\theta_\rho)\, e^{i\sigma_\tau\phi_\rho}\,.
\label{eq:tau_rho_helicity_amplitude}
\end{eqnarray}
Note that the decay with the $\rho^-$ helicity of $\lambda=+1$ is forbidden due
to the angular momentum conservation. Folding the polarized $\tau^-$ decay
distributions with a given $\tau^-$ polarization matrix and integrating them
over the azimuthal angle $\phi_\rho$ yield the polar-angle dependent
distributions in the $(0,-1)$ basis
\begin{eqnarray}
  W_{\sigma\lambda}(\theta_\rho)
=\left[\begin{array}{cc}
       \frac{m^2_\tau}{m^2_\tau+2m^2_\rho}(1+P_\tau \cos\theta_\rho)
    & -\frac{\sqrt{2} m_\tau m_\rho}{m^2_\tau+2m^2_\rho}P_\tau\sin\theta_\rho
    \\[4mm]
      -\frac{\sqrt{2} m_\tau m_\rho}{m^2_\tau+2m^2_\rho}P_\tau\sin\theta_\rho
    & \frac{2m^2_\rho}{m^2_\tau+2m^2_\rho}(1-P_\tau \cos\theta_\rho)
        \end{array}\right]\,,
\label{eq:rho_density_matrix_tau_rest_frame}
\end{eqnarray}
apart from an overall factor. The average of the diagonal elements is the
normalized polar-angle distribution of $\rho^-$ in the $\tau^-$ rest frame
\begin{eqnarray}
  \overline{W}(\theta_\rho)
= \frac{1}{2}\left[ 1
                   +\left(\frac{m^2_\tau-2m^2_\rho}{m^2_\tau+2m^2_\rho}\right)\,
                    P_\tau \cos\theta_\rho\right]\,.
\end{eqnarray}
The polarization-dependent distribution matrix in
Eq.$\,$(\ref{eq:rho_density_matrix_tau_rest_frame}) cannot be directly used
before being combined with the $\rho$ decay part, but it first must be
transformed by the Wick helicity rotation into the corresponding
polarization-dependent distribution in the $e^-e^+$ CM frame, i.e. the LAB
frame as
\begin{eqnarray}
  W'_{\sigma'\lambda'}(\omega,\theta_\rho)
= \sum_{\sigma,\lambda} d^1_{\sigma'\sigma}(\omega) d^1_{\lambda'\lambda}(\omega)
   \, W_{\sigma\lambda}(\theta_\rho)\,.
\end{eqnarray}
Although it is straightforward to derive the full expression of the
distribution matrix in the LAB, we restrict ourselves to the diagonal elements,
since we consider only the polar-angle distributions and a parity-conserving
decay $\rho^-\to\pi^-\pi^0$. An explicit evaluation leads to the following
transverse and longitudinal distributions of the $\rho^-$ meson,
\begin{eqnarray}
&& W'_T(\omega,\theta_\rho)
 = \frac{m^2_\tau+m^2_\rho}{m^2_\tau+2m^2_\rho}
   -\frac{m^2_\tau-m^2_\rho}{m^2_\tau+2m^2_\rho}\cos^2\omega \nonumber\\
&&\hskip 1.5cm +\,P_\tau \left[ \left(\frac{m^2_\tau-m^2_\rho}{m^2_\tau+2m^2_\rho}
   -\frac{m^2_\tau+m^2_\rho}{m^2_\tau+2m^2_\rho}\cos^2\omega\right)\cos\theta_\rho
   -\frac{2m_\tau m_\rho}{m^2_\tau+2m^2_\rho}\cos\omega\sin\omega\sin\theta_\rho
   \right]\,,
   \label{eq:transverse_rho_polarization}\\
&&  W'_L(\omega,\theta_\rho)
 = \frac{m^2_\rho}{m^2_\tau+2m^2_\rho}
   +\frac{m^2_\tau-m^2_\rho}{m^2_\tau+2m^2_\rho}\cos^2\omega \nonumber\\
&&\hskip 1.5cm -\,P_\tau \left[ \left(\frac{m^2_\rho}{m^2_\tau+2m^2_\rho}
   -\frac{m^2_\tau+m^2_\rho}{m^2_\tau+2m^2_\rho}\cos^2\omega\right)\cos\theta_\rho
   -\frac{2m_\tau m_\rho}{m^2_\tau+2m^2_\rho}\cos\omega\sin\omega\sin\theta_\rho
   \right]\,,
   \label{eq:longitudinal_rho_polarization}
\end{eqnarray}
in the LAB frame to be folded {\it directly} with the $\rho^-$ decay
distributions in the
$\rho^-$ rest frame.\\

\begin{figure}[htb]
\centering
\includegraphics[width=17cm]{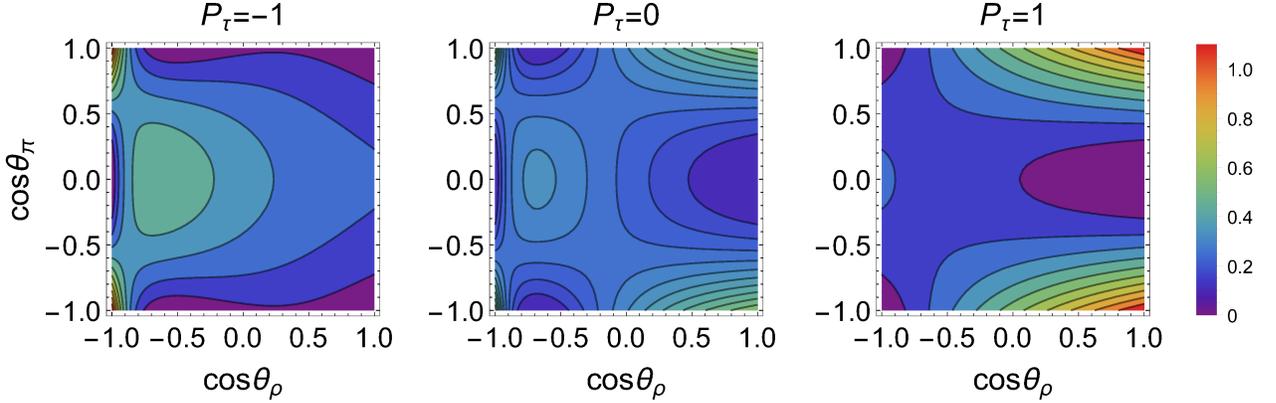}
\caption{\it Contour plots of the polar-angle correlation function
             $d^2{\cal C}(\theta_\rho,\theta_\pi)/d\cos\theta_\rho\, d\cos\theta_\pi$
             for $P_\tau=-1$ (left), $P_\tau=0$ (middle) and $P_\tau=1$ (right)
             of the decaying $\tau$ longitudinal polarization. Here, the $e^+e^-$
             CM energy is set to the $Z$ mass $m_Z=91\, {\rm GeV}$
             and $m_\tau=1.78\, {\rm GeV}$
             and $m_\rho=0.77\, {\rm GeV}$ are chosen.
             }
\label{fig:correlation_function_tau_rho_pi}
\end{figure}

The $\rho^-$ decays via $\rho^-\to\pi^-\pi^0$ with almost 100\% probability. By
the conserved vector current (CVC) hypothesis
\cite{Gershtein:1955fb,Feynman:1958ty}, the $\rho^-$ decay mode can be
completely described by the four-vector current as
\begin{eqnarray}
  D[\rho^-\to\pi^-\pi^0]
= f_\rho (p_{\pi^-}-p_{\pi^0})_\mu \epsilon^\mu\,,
\end{eqnarray}
where $\epsilon^\mu$ is the $\rho^-$ polarization vector. The helicity
amplitudes can be cast into the simple form in the helicity basis
\begin{eqnarray}
  D_{\sigma_\rho}(\theta_\pi,\phi_\pi)
\ \ \sim \ \
  d^1_{\sigma_\rho,0}(\theta_\pi)\, e^{i\sigma_\rho\phi_\pi}\,,
\label{eq:rho_pipi_helicity_amplitude}
\end{eqnarray}
in the $\rho^-$ rest frame, leading to the decay angular distributions
\cite{Hagiwara:1989fn} for the transversely and longitudinally polarized
$\rho^-$
\begin{eqnarray}
&& \frac{1}{\Gamma}\,\frac{d\Gamma(\rho_T\to 2\pi)}{d\cos\theta_\pi}
   = \frac{3}{8}\,\sin^2\theta_\pi\,,
   \label{eq:transverse_rho_pipi_distribution} \\[2mm]
&& \frac{1}{\Gamma}\, \frac{d\Gamma(\rho_L\to 2\pi)}{d\cos\theta_\pi}
   = \frac{3}{4}\,\cos^2\theta_\pi\,,
   \label{eq:longitudinal_rho_pipi_distribution}
\end{eqnarray}
with $\cos\theta_\pi=(2E_\pi/E_\rho-1)/\sqrt{1-4m^2_\pi/m^2_\rho}$ in the
collinear limit.\\

\begin{figure}[htb]
\centering
\includegraphics[width=12cm, height=9cm]{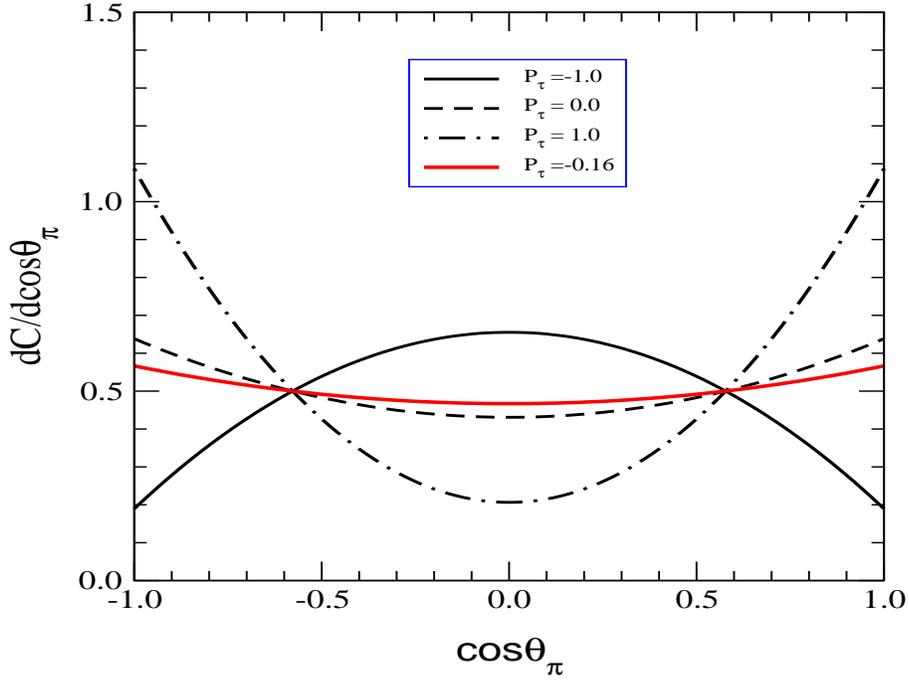}
\caption{\it Polar angle distribution $d{\cal C}/d\cos\theta_\pi$ in the two-stage
             decay $\tau^-\to\rho^-\nu_\tau\to (\pi^-\pi^0)\nu_\tau$
             for the degrees of $\tau$ polarization of $P_\tau=\pm 1, 0$ (taken simply
             for comparison) and $P_\tau=-0.16$, the value expected with the SM
             couplings.
             }
\label{fig:pi_polar-angle_distribution}
\end{figure}

Eventually, combining Eqs.$\,$(\ref{eq:transverse_rho_polarization}) and
(\ref{eq:longitudinal_rho_polarization}) with
Eqs.$\,$(\ref{eq:transverse_rho_pipi_distribution}) and
(\ref{eq:longitudinal_rho_pipi_distribution}) we can obtain the full normalized
spin and polar-angle correlations of the two-stage decays
$\tau^-\to\rho^-\nu_\tau \to (\pi^-\pi^0)\nu_\tau$:
\begin{eqnarray}
  \frac{d^2{\cal C}}{d\cos\theta_\rho\, d\cos\theta_\pi}
= \frac{1}{4}\bigg[\,f(\theta_\rho,\theta_\pi)
  + P_\tau\, g(\theta_\rho,\theta_\pi)\,\bigg]
\,,
\label{eq:correlation_function_rho_pi}
\end{eqnarray}
with the functions containing the $(\theta_\rho, \theta_\pi)$ correlations
given by
\begin{eqnarray}
&& f(\theta_\rho, \theta_\pi)
 = 1+\frac{1}{2}\frac{m^2_\tau-m^2_\rho}{m^2_\tau+2m^2_\rho}\,
               (3\cos^2\omega-1)(3\cos^2\theta_\pi-1)\,, \\
&& g(\theta_\rho, \theta_\pi)
 = \frac{m^2_\tau-2m^2_\rho}{m^2_\tau+2m^2_\rho}\, \cos\theta_\rho
        +\frac{1}{2}\frac{m^2_\tau+m^2_\rho}{m^2_\tau+2m^2_\rho}
         (3\cos^2\omega-1)\cos\theta_\rho\, (3\cos^2\theta_\pi-1)
         \nonumber\\
&& \hskip 1.2cm +3\frac{m_\tau m_\rho}{m^2_\tau+2m^2_\rho}
    \cos\omega \sin\omega \sin\theta_\rho\, (3\cos^2\theta_\pi-1)\,,
\end{eqnarray}
which are consistent with the corresponding expressions in
Refs.~\cite{Hagiwara:1989fn,Bullock:1992yt}. We note that the Wick helicity
rotation angle $\omega$ is a function of $\theta_\rho$, the polar angle of
$\rho^-$ in the $\tau^-$ rest frame and it depends on $\beta_\tau$ in the LAB
and $\beta_\rho$ fixed with the $\tau$ and $\rho$ masses. Thus $P_\tau$ can be
determined from an analysis of the two-dimensional distribution with greatly
improved precision as demonstrated numerically in
Fig.$\,$\ref{fig:correlation_function_tau_rho_pi}.\\

If the correlation function in Eq.$\,$(\ref{eq:correlation_function_rho_pi}) is
integrated over the polar angle $\theta_\rho$ of the $\rho^-$, then we obtain
the single polar angle distribution $d{\cal C}/d\cos\theta_\pi$ of $\pi^-$
expressed in terms of PEFs described in detail in Appendix~
\ref{appendix:polarization_estimator_functions}. In this case with
$\beta_\tau\simeq 1$, the asymptotic expressions of PEFs can be safely used. We
note that the polar-angle distribution, symmetric due to the parity-preserving
decay $\rho^-\to\pi^-\pi^0$, is quite sensitive to the value of $P_\tau$ as
clearly indicated by Fig.$\,$\ref{fig:pi_polar-angle_distribution}.
Nevertheless, it is certain that the full polar-angle correlation enables us to
determine $P_\tau$
and so the weak-mixing angle $\theta_W$ with greater precision.\\

\subsection{The process \boldmath{$e^-e^+\to t\,\bar{t}$} followed by
            \boldmath{$t\to W^+ b$} and
            \boldmath{$W^+\to \ell^+\nu_\ell$}}
\label{subsec:top_pair_production}

Studying top quarks with great precision after its discovery at Tevatron
\cite{Abe:1995hr,D0:1995jca} is important in particular for several theoretical
and experimental reasons. It allows us to probe physics at a much higher mass
scale than the other SM fermions. To a very good approximation the top quark
decays as a free quark, because of the top quark lifetime of about $4.7\times
10^{-25}\,{\rm s}$ (corresponding to the width of $1.41\, {\rm GeV}$) is too
short for the top quark to bind with light quarks before it decays
\cite{Bigi:1986jk}. Furthermore, the maximally parity-violating two-body
top-quark decay $t\to W^+ b$ enables us to analyze the top-quark polarization
efficiently, which is in general non-zero if its production proceeds through
some parity-violating interactions
\cite{Shelton:2008nq,V.:2016wba,Kane:1991bg,Ladinsky:1992vv,Choi:1995kp,
Mahlon:1995zn,Parke:1996pr,Asakawa:2000jy,
BhupalDev:2007ftb,Bernreuther:2008ju,Choudhury:2010cd,Godbole:2010kr,
Prasath:2014mfa,Tweedie:2014yda,Godbole:2015bda}.\\

In this subsection, as another characteristic example of the spin combination
$1/2\to 1 +1/2$, we consider the following three-stage top production and decay
processes at the tree level in the SM:
\begin{eqnarray}
e^-e^+ \to t\bar{t} \to (W^+b)\, \bar{t} \to ([\ell^+\nu_\ell]\,b)\, \bar{t}\,,
\end{eqnarray}
including the two-body decay $t\to W^+ b$ as the key chain, with $\ell=e, \mu$,
while treating $\bar{t}$ inclusively.\\

The $t$-pair production in $e^-e^+$ annihilation proceeds via two $s$-channel
$\gamma$ and $Z$ exchanges. The Feynman rules of the $ee V$ and $V tt$
couplings with $V=\gamma, Z$  are
\begin{eqnarray}
    \langle \gamma^\mu | e^-e^+\rangle
&=& ie\,\gamma^\mu\,, \\
    \langle t\bar{t}| \gamma_\mu \rangle
&=& -ie Q_t \gamma_\mu\,, \\
    \langle Z^\mu | e^-e^+\rangle
&=& ig_Z\,\gamma^\mu (v_e-a_e\gamma_5)\,,\\
    \langle t\bar{t}| Z_\mu \rangle
&=& ig_Z \gamma_\mu\, (v_t-a_t\gamma_5)\,,
\end{eqnarray}
with the normalized $t$ electric charge $Q_t=2/3$ and the vector and axial-vector
couplings $v_t= 1/4 -2 s^2_W/3$ and $a_t= 1/4$. \\

By introducing four bilinear charges \cite{Sehgal:1980ja,Choi:2000ta} defined
by
\begin{eqnarray}
Q_{\pm\pm} &=& Q_t - \frac{\gamma_Z(s)}{c^2_W s^2_W} (v_e\mp a_e) (v_t\mp a_t)\,, \\
Q_{\pm\mp} &=& Q_t - \frac{\gamma_Z(s)}{c^2_W s^2_W} (v_e\mp a_e) (v_t\pm a_t)\,,
\end{eqnarray}
with $\gamma_Z(s)=s/(s-m^2_Z+im_Z\Gamma_Z)$, the helicity amplitudes in the $e^-e^+$
CM frame can be written in a compact form as
\begin{eqnarray}
  T(\sigma\bar{\sigma}: \lambda\bar{\lambda})
= - 2\pi\alpha\, \delta_{\bar{\sigma},-\sigma}\,
  \langle \sigma: \lambda\bar{\lambda}\rangle\,
  e^{i(\sigma-\bar{\sigma})\Phi}\,,
\end{eqnarray}
with the replacement $e^2=4\pi\alpha$ by the fine-structure constant $\alpha$
and with $m_e=0$ assumed, where the helicity-dependent parts \cite{Kane:1991bg}
are
\begin{eqnarray}
  \langle \pm: ++\rangle
&=& \mp (1-\beta^2_t)^{1/2} (Q_{\pm +} + Q_{\pm -}) \sin\Theta\,, \\[2mm]
 \langle \pm: +-\rangle
&=& [(1+\beta_t) Q_{\pm +} + (1-\beta_t) Q_{\pm -}] (1\pm \cos\Theta)\,, \\[2mm]
 \langle \pm: -+\rangle
&=& [(1-\beta_t) Q_{\pm +} + (1+\beta_t) Q_{\pm -}] (1\mp \cos\Theta)\,, \\[2mm]
   \langle \pm: --\rangle
&=& \mp (1-\beta^2_t)^{1/2} (Q_{\pm +} + Q_{\pm -}) \sin\Theta\,.
\end{eqnarray}
We note that the non-zero $Z$ width can in general be neglected for the
energies considered in the present analysis so that the bilinear charges are
real with very good approximation at the tree level.\\

For the sake of notational convenience we introduce six quartic charges
\cite{Sehgal:1980ja,Choi:2000ta} for the $t$-pair production process. These
charges correspond to independent helicity-dependent components describing the
$t$-pair production for polarized electrons and positrons with negligible
electron mass. Three parity-even (unprimed) quartic charges are defined in
terms of bilinear charges as
\begin{eqnarray}
Q_1 &=& \frac{1}{4}\left( |Q_{++}|^2 +|Q_{+-}|^2+|Q_{-+}|^2+|Q_{--}|^2\right)\,,
  \\[1mm]
Q_2 &=& \frac{1}{2} {\rm Re}\left(Q_{++}Q^*_{+-}+Q_{--}Q^*_{-+}\right)\,,\\[1mm]
Q_3 &=& \frac{1}{4}\left( |Q_{++}|^2 -|Q_{+-}|^2-|Q_{-+}|^2+|Q_{--}|^2\right)\,,
\end{eqnarray}
and three parity-odd (primed) quartic charges as
\begin{eqnarray}
Q'_1 &=& \frac{1}{4}\left( |Q_{++}|^2 +|Q_{+-}|^2-|Q_{-+}|^2-|Q_{--}|^2\right)\,,
  \\[1mm]
Q'_2 &=& \frac{1}{2} {\rm Re}\left(Q_{++}Q^*_{+-}-Q_{--}Q^*_{-+}\right)\,,\\[1mm]
Q'_3 &=& \frac{1}{4}\left( |Q_{++}|^2 -|Q_{+-}|^2+|Q_{-+}|^2-|Q_{--}|^2\right)\,.
\end{eqnarray}
In terms of these six quartic charges, the differential cross section for
longitudinally polarized electron and positron beams and the degree of
longitudinal $t$ polarization are given in a simple form by
\begin{eqnarray}
    \frac{d\sigma_{\rm pol}}{d\cos\Theta}
&=& \frac{\pi\alpha^2}{2s}\beta_t
    \left[ (1-P_{e^-}P_{e^+}) \Sigma_{\rm unp}
          +(P_{e^-}-P_{e^+}) \Sigma_{LL}\right)\,,
\label{eq:tt-bar_differential_x_section}\\
    P_t(\Theta)
&=& \frac{(1-P_{e^-}P_{e^+}) \Delta_{\rm unp}+(P_{e^-}-P_{e^+}) \Delta_{LL}}{
          (1-P_{e^-}P_{e^+}) \Sigma_{\rm unp}+(P_{e^-}-P_{e^+}) \Sigma_{LL}}\,.
\label{eq:angular_dependent_t_polarization}
\end{eqnarray}
The polar-angle dependent coefficients, $\Sigma_{\rm unp}$, $\Sigma_{LL}$,
$\Delta_{\rm unp}$ and $\Delta_{LL}$, appearing in
Eqs.$\,$(\ref{eq:tt-bar_differential_x_section}) and
(\ref{eq:angular_dependent_t_polarization}) are expressed in terms of the
quartic charges as
\begin{eqnarray}
    \Sigma_{\rm unp}
&=& (1+\beta^2_t\cos^2\Theta)\, Q_1 + (1-\beta^2_t)\, Q_2 + 2\beta_t\, Q_3 \cos\Theta\,,
\\[2mm]
    \Sigma_{LL}
&=& (1+\beta^2_t\cos^2\Theta)\, Q'_1 + (1-\beta^2_t)\, Q'_2 + 2\beta_t\, Q'_3 \cos\Theta\,,
\\[2mm]
    \Delta_{\rm unp}
&=& [(1+\beta^2_t)\, Q'_1 + (1-\beta^2_t)\, Q'_2]\cos\Theta + \beta_t\, Q'_3 (1+\cos^2\Theta)\,,
\\[2mm]
   \Delta_{LL}
&=& [(1+\beta^2_t)\, Q_1 + (1-\beta^2_t)\, Q_2]\cos\Theta + \beta_t\, Q_3 (1+\cos^2\Theta)\,.
\end{eqnarray}
If the production angles could be measured unambiguously on an event-by-event basis, the
quartic charges could be extracted directly from the angular dependence of the
cross section equipped with polarized electron and/or positron beams at a single
energy and similarly from the direct measurement of longitudinal $t$ polarization.
However, the (longitudinal) $t$ polarization can only be determined indirectly from
angular distribution of decay products if the $t$ decay dynamics is known.\\

The top quark $t$ with its mass of about 173 GeV decays via the
parity-violating weak decay $t\to W^+b$ with almost 100\% probability
\cite{Tanabashi:2018oca}, of which the dynamical structure is identical to that
of $\tau^-\to\rho^-\nu_\tau$. The decay mode with $W^+$ of helicity $+1$ in the
$t$ rest frame is forbidden because of angular momentum conservation. Thus
folding the polarized $t$ decay distributions with a given $t$ polarization
matrix and integrating the resulting distributions over the $W^+$ azimuthal
angle yield the polar-angle dependent distributions in the $(0,-1)$ basis of
$W^+$, while ignoring the $+1$ mode with vanishing components, as
\begin{eqnarray}
  W_{\sigma\lambda}(\theta_W)
=\left[\begin{array}{cc}
       \frac{m^2_t}{m^2_t+2m^2_W}(1+P_t \cos\theta_W)
    & -\frac{\sqrt{2} m_t m_W}{m^2_t+2m^2_W}P_t\sin\theta_W \\[3mm]
      -\frac{\sqrt{2} m_t m_W}{m^2_t+2m^2_W}P_t\sin\theta_W
    & \frac{2m^2_W}{m^2_t+2m^2_W}(1-P_t \cos\theta_W)
        \end{array}\right]\,,
\label{eq:w_density_matrix_t_rest_frame}
\end{eqnarray}
of which the average of the diagonal elements leads to the normalized polar-angle
distribution of $W^+$ in the $t$ rest frame:
\begin{eqnarray}
  \overline{W}(\theta_W)
= \frac{1}{2}\left[ 1
                   +\left(\frac{m^2_t-2m^2_W}{m^2_t+2m^2_W}\right)\,
                    P_t\, \cos\theta_W\right]\,.
\end{eqnarray}
In order to connect the $W^+$ polarization density matrix in the $t$ rest frame
directly with the $W^+$ decay distribution in the $W^+$ rest frame, it is
necessary to transform the density matrix in
Eq.$\,$(\ref{eq:w_density_matrix_t_rest_frame}) by Wick helicity rotation.
Although it is straightforward to derive the full expression of the matrix in
the $e^-e^+$ CM frame, we will restrict ourselves to the derivation of the
diagonal elements. An explicit calculation leads to the following distributions
\begin{eqnarray}
 W'_{\pm\pm}(\omega,\theta_W)
&=& \frac{1}{2}\left[W'_{T}(\omega,\theta_W)\pm Z'_T(\omega,\theta_W)\right]\,,
\label{eq:w_pm_distribution}   \\[2mm]
 W'_{00}(\omega,\theta_W)
&=&  W'_L(\omega,\theta_W)\,,
\label{eq:w_00_distribution}
\end{eqnarray}
with the parity-even and parity-odd transverse parts, $W'_T$ and $Z'_T$, given
explicitly by
\begin{eqnarray}
&& W'_T(\omega,\theta_W)
 = \frac{m^2_t+m^2_W}{m^2_t+2m^2_W}
   -\frac{m^2_t-m^2_W}{m^2_t+2m^2_W}\cos^2\omega \nonumber\\
&&\hskip 1.5cm +\,P_t \left[ \left(\frac{m^2_t+m^2_W}{m^2_t+2m^2_W}
   -\frac{m^2_t+m^2_W}{m^2_t+2m^2_W}\cos^2\omega\right)\cos\theta_W
   -\frac{2m_t m_W}{m^2_t+2m^2_W}\cos\omega\sin\omega\sin\theta_W\right]\,,
   \label{eq:transverse_w_polarization_sum}\\
&& Z'_T(\omega,\theta_W)
 = -\frac{2m^2_W}{m^2_t+2m^2_W} \cos\omega
   + 2P_t \left(\frac{m^2_W}{m^2_t+2m^2_W}\cos\omega\cos\theta_W
                 +\frac{m_t m_W}{m^2_t+2m^2_W}\sin\omega\sin\theta_W\right)\,,
  \label{eq:transverse_w_polarization_difference}
\end{eqnarray}
and the parity-even longitudinal part $W'_L$ given explicitly by
\begin{eqnarray}
&& W'_L(\omega,\theta_W)
 = \frac{m^2_W}{m^2_t+2m^2_W}
   +\frac{m^2_t-m^2_W}{m^2_t+2m^2_W}\cos^2\omega \nonumber\\
&&\hskip 1.5cm -\,P_t \left[ \left(\frac{m^2_W}{m^2_t+2m^2_W}
   -\frac{m^2_t+m^2_W}{m^2_t+2m^2_W}\cos^2\omega\right)\cos\theta_W
   -\frac{2m_t m_W}{m^2_t+2m^2_W}\cos\omega\sin\omega\sin\theta_W\right]\,,
   \label{eq:longitudinal_w_polarization}
\end{eqnarray}
in the LAB to be folded with the $W^+$ decay distributions in the
$W^+$ rest frame boosted directly back from the LAB.\\

\begin{figure}[htb]
\centering
\includegraphics[width=17.cm]{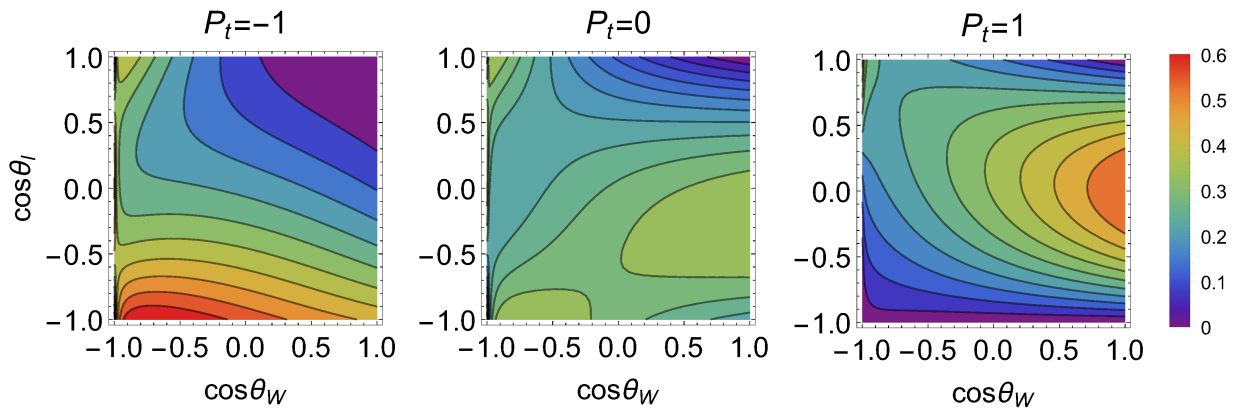}
\caption{\it Contour plots of the polar-angle correlation function
             $d^2{\cal C}/d\cos\theta_W d\cos\theta_\ell$ for $P_t=1$ (left),
             $P_t=0$ (middle), and $P_t=-1$ (right) values of the decaying
             $t$ longitudinal polarization. Here, the $e^+e^-$
             CM energy of $500\, {\rm GeV}$ is taken and $m_t=173\, {\rm GeV}$
             and $m_W=80.4\, {\rm GeV}$ are assumed.
             }
\label{fig:correlation_function_twl}
\end{figure}

The $W^+$ weak decay into a positive lepton $\ell^+$ and its neutrino
$\nu_\ell$ with $\ell=e$ or $\mu$, accounting for the branching fraction of
about 20\%, is a very clean signal for diagnosing the $W^+$ polarization.
Neglecting the lepton mass with good approximation, i.e. setting $m_\ell=0$, we
can obtain the decay helicity amplitudes in the $W^+$ rest frame as
\begin{eqnarray}
D_{\sigma_W} (\theta_\ell,\phi_\ell)
 = \frac{g}{\sqrt{2}}\bar{u}(k,-) \gamma^\mu P_L v(q,+)
                  \epsilon_\mu(p,\sigma)
                  = gm_W\, d^1_{\sigma_W, 1}(\theta_\ell)
                  \, e^{i\sigma_W\phi_\ell}\,,
\end{eqnarray}
leading to the normalized amplitudes
\begin{eqnarray}
&& D_\pm (\theta_\ell,\phi_\ell)
  = \frac{1\pm\cos\theta_\ell}{2}\, e^{\pm i \phi_\ell}\,,
   \label{eq:w_pm_decay_distribution} \\
&& D_0 (\theta_\ell, \phi_\ell)
  = \frac{\sin\theta_\ell}{\sqrt{2}}\,,
\label{eq:w_00_decay_distribution}
\end{eqnarray}
satisfying $|D_+|^2+|D_0|^2+|D_-|^2=1$. Combining
Eqs.$\,$(\ref{eq:w_pm_distribution}) and (\ref{eq:w_00_distribution}) with
Eqs.$\,$(\ref{eq:w_pm_decay_distribution}) and
(\ref{eq:w_00_decay_distribution}) yields the full spin and polar-angle
correlations of the two-stage decays $t\to W^+b\to (\ell^+\nu_\ell) b$:
\begin{eqnarray}
  \frac{d^2{\cal C}}{d\cos\theta_W d\cos\theta_\ell}
= \frac{1}{4} \bigg[\,f(\theta_W,\theta_\ell)
  +P_t\, g(\theta_W,\theta_\ell)\,\bigg]\,,
\label{eq:correlation_function_twl}
\end{eqnarray}
with the two $\theta_W$ and $\theta_\ell$ correlation functions of which the
first function
\begin{eqnarray}
f(\theta_W,\theta_\ell)
 = 1-\frac{3m^2_W}{m^2_t+2m^2_W}\cos\omega\cos\theta_\ell
  -\frac{1}{4}\frac{m^2_t-m^2_W}{m^2_t+2m^2_W}(3\cos^2\omega-1)(3\cos^2\theta_\ell-1)
 \,,
\end{eqnarray}
surviving even for unpolarized $t$ and the second function
\begin{eqnarray}
&& g(\theta_W,\theta_\ell)
 = \frac{m^2_t-2m^2_W}{m^2_t+2m^2_W}\cos\theta_W
   +\frac{3m^2_W}{m^2_t+2m^2_W}\cos\omega\cos\theta_W\cos\theta_\ell\nonumber\\
&& \hskip 1.5cm
   -\frac{1}{4}\frac{m^2_t+m^2_W}{m^2_t+2m^2_W}(3\cos^2\omega-1)\cos\theta_W
     (3\cos^2\theta_\ell-1)\nonumber\\
&& \hskip 1.5cm
   +\frac{3m_t m_W}{m^2_t+2m^2_W}\sin\omega\sin\theta_W\cos\theta_\ell
   -\frac{1}{2}\frac{3m_t m_W}{m^2_t+2m^2_W}
   \cos\omega\sin\omega \sin\theta_W (3\cos^2\theta_\ell-1)\,,
\end{eqnarray}
contributing only when the $t$ quark is polarized. As mentioned before, the
Wick helicity rotation angle $\omega$ is a function of $\theta_W$, the polar
angle of $W^+$ in the $t$ rest frame, and two boost factors, $\beta_t$ and
$\beta_W=(m^2_t-m^2_W)/(m^2_t+m^2_W)$. Thus $P_t$ can be determined efficiently
from an analysis of the two-dimensional angular distribution as demonstrated
clearly with three values of $P_t=\pm 1, 0$ (for the sake of simple comparison)
at the $e^-e^+$ CM energy of $500\, {\rm GeV}$ in
Fig.$\,$\ref{fig:correlation_function_twl}.\\

\begin{figure}[htb]
\centering
\includegraphics[width=12cm, height=9cm]{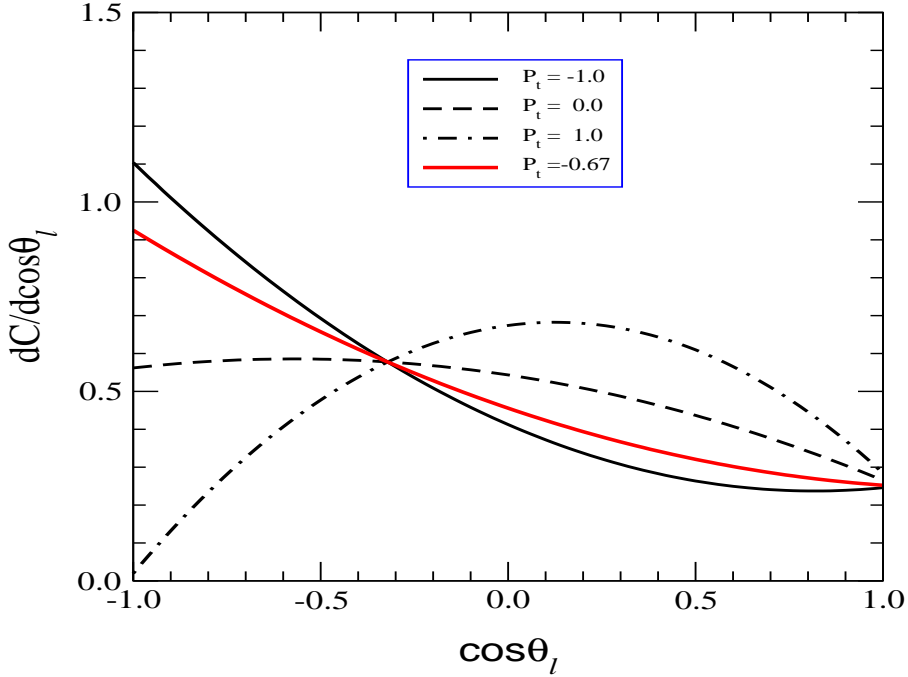}
\caption{\it Polar angle distribution $d{\cal C}/d\cos\theta_\ell$ in the two-stage
             decay $t\to W^+ b\to (\ell^+\nu_\ell)t$
             for the degrees of $\tau$ polarization of $P_t=\pm 1, 0$ for the sake
             of simple comparison and
             $P_t=-0.67$, the value expected with the SM $\gamma$ and $Z$ couplings
             to a pair of top quarks.
             }
\label{fig:l_single_polar-angle_distribution}
\end{figure}

If the correlation function in Eq.$\,$(\ref{eq:correlation_function_twl}) is
integrated over the polar angle $\theta_W$ of the $W^+$, then we can obtain the
single polar angle distribution $d{\cal C}/d\cos\theta_\ell$ of $\ell^-$
expressed in terms of polarization estimators described in detail in
appendix~\ref{appendix:polarization_estimator_functions}. We note that the
polar-angle distribution, asymmetric due to the parity-violating decay
$W^+\to\ell^+\nu_\ell$, is quite sensitive to the value of $P_t$ as shown in
Fig.$\,$\ref{fig:l_single_polar-angle_distribution}. Nevertheless, as mentioned
before, it is certain that the full polar-angle correlation enables us to
determine $P_\tau$ and so
the weak-mixing angle $\theta_W$ with better precision.\\

\subsection{The decay \boldmath{$T\to Z t$} of a heavy vectorlike top quark $T$,
            followed by \boldmath{$Z\to\ell^-\ell^+$}}
\label{subsec:T_Zt_llt}

In many models of new physics beyond the SM such as extra-dimensional models
and little Higgs models
\cite{Perelstein:2003wd,Cacciapaglia:2010vn,Yue:2007py,Han:2008xb,Cheng:2005as,
Schmaltz:2005ky,Agashe:2004rs,Schmaltz:2004de,Han:2003wu,Zhou:2019alr}, there
are heavy vectorlike fermions which decay to the SM fermions plus a gauge boson
($W^\pm$ and $Z$) or a Higgs boson ($H$). The mixing of vector-like quarks with
the third generation and in particular with the top quark is a common feature
in little Higgs models and it may be
sizable.\\

Due to its heavy mass, the new colored vectorlike heavy fermion $T$ may only be
produced at high energy hadron colliders. The apparent production processes are
the QCD pair production, $q\bar{q}, gg \to T\bar{T}$, producing unpolarized $T$
particles. However, the phase space suppression for the heavy TeV-scale mass is
rather severe in the pair production. In contrast, the single $T$ production
via $W$ exchange in $t$-channel (or $Wb$ fusion) $qb \to T q'$, in which the
particle $T$ is produced in a polarized state, falls off much more slowly with
the $T$ mass and takes over for $m_T$ larger than a few hundred GeV. According
to the so-called Goldstone boson equivalence theorem
\cite{Cornwall:1974km,Vayonakis:1976vz}, the $T$ couplings to the
longitudinally polarized gauge bosons are not suppressed, rendering the decay
$T\to Z t$ being one of the main decay channels. The $Z$ boson in the final
state gives an unambiguous event identification via its clean leptonic decay,
and the system $t(\to Wb)Z$
enables us to reconstruct $m_T$ \cite{Han:2003wu}. \\

Without taking any specific model into account, we assume a generic chiral
structure for the $TtZ$ interaction vertex of a heavy and SM top quarks, $T$
and $t$, and the neutral gauge boson $Z$, denoting the vector and axial-vector
couplings by $v_T$ and $a_T$ normalized with the SM gauge coupling $g_Z=e/c_W
s_W$ as
\begin{eqnarray}
    \langle t| Z | T\rangle_\mu
= -ig_Z\,\gamma_\mu (v_T-a_T\gamma_5)\,.
\end{eqnarray}
The helicity amplitude of the two-body decay $T\to Zt$ with its expected
branching fraction larger than 20\% is written in the $T$ rest frame as
\begin{eqnarray}
   D_{\sigma_T; \lambda, \sigma_t}(\theta_Z,\phi_Z)
= F^{1/2}_{\lambda\sigma_t}\, d^{1/2}_{\sigma_T, \lambda-\sigma_t}(\theta_Z)\,\,
  e^{i\sigma_T\phi_Z}\,,
\end{eqnarray}
where $\theta_Z$ and $\phi_Z$ are the polar and azimuthal angles of the $Z$ boson.
Apart from an overall factor, the angle-independent reduced helicity amplitudes
read
\begin{eqnarray}
F^{1/2}_{\pm \pm} = \sqrt{2}\, (v_1\mp a_1) \ \ \mbox{and} \ \
F^{1/2}_{0\pm}    = \sqrt{2}\, (v_0\mp a_0)\,,
\end{eqnarray}
in terms of the redefined vector and axial-vector couplings as
\begin{eqnarray}
&& v_1 = \sqrt{(1-\mu_t)^2-\mu^2_Z}\, v_T\,,\qquad\quad\quad\quad\ \
   a_1 = \sqrt{(1+\mu_t)^2-\mu^2_Z}\, a_T \,, \\
&& v_0 = \frac{(1+\mu_t)}{\sqrt{2}\mu_Z} \sqrt{(1-\mu_t)^2-\mu^2_Z}\, v_T\,,
   \quad\quad
   a_0 = \frac{(1-\mu_t)}{\sqrt{2}\mu_Z} \sqrt{(1+\mu_t)^2-\mu^2_Z}\, a_T \,,
\end{eqnarray}
with the normalized dimensionless mass ratios, $\mu_t=m_t/m_T$ and $\mu_Z = m_Z/m_T$.\\

Integrating the decay distribution derived from the helicity amplitudes over
the azimuthal angle $\phi_Z$ and folding with the $T$ polarization $P^T$ yield
the helicity-dependent distributions\footnote{The reason why ${\cal
D}_{\pm\mp}=0$ is due to angular momentum conservation.}
\begin{eqnarray}
&&    {\cal D}_{\pm\pm}(\theta_Z)
\,=\, |v_1\mp a_1|^2 (1\pm P^T\cos\theta_Z)\,, \\[2mm]
&&    {\cal D}_{\pm\mp}(\theta_Z)
\,= \,0\,, \\[2mm]
&&    {\cal D}_{+\,0}(\theta_Z)
\,=\, {\cal D}^*_{0\,+}\, =\, -(v_1-a_1)(v_0-a_0)^* P^T \sin\theta_Z\,, \\[2mm]
&&    {\cal D}_{-\,0}(\theta_Z)
\,=\, {\cal D}^*_{0\,-}\, =\, -(v_1+a_1) (v_0+a_0)^* P^T \sin\theta_Z\,, \\[2mm]
&&    {\cal D}_{\,0\,0}(\theta_Z)
\,=\, 2(|v_0|^2 + |a_0|^2) + 4{\rm Re}(v_0 a^*_0) P^T \cos\theta_Z\,,
\end{eqnarray}
apart from an overall factor. The average of the diagonal elements leads to the
normalized polar-angle distribution in the $T$ rest frame:
\begin{eqnarray}
  \overline{W}(\theta_Z)
= \frac{1}{2}\left(1+AP^T\cos\theta_Z\right)\quad
\mbox{with}\quad
A= \frac{2\,{\rm Re}\,(v_0 a^*_0-v_1 a^*_1)}{|v_1|^2+|a_1|^2+|v_0|^2+|a_0|^2}\,.
\end{eqnarray}
In order to connect the $Z$ polarization density matrix in the $T$ rest frame
{\it directly} with the $Z$ decay distribution in the LAB we transform it into
the $Z$ density matrix in the LAB by Wick helicity rotation. Although it is
straightforward to derive the full expression in any given LAB, in the present
work we restrict ourselves to the derivation of the diagonal elements for a
fixed $T$ speed, $\beta_T$. As the transformed distributions involve various
combinations of the redefined couplings, $\{v_1, a_1, v_0, a_0\}$, let us first
introduce five ratios consisting of three parity-odd ratios
\begin{eqnarray}
A_1 &=& - \frac{2\,{\rm Re}(v_1 a^*_1)}{|v_1|^2+|a_1|^2+|v_0|^+|a_0|^2}\,, \\
A_2 &=& -\frac{2\,{\rm Re}(v_1 a^*_0 + a_1 v^*_0)}{|v_1|^2+|a_1|^2+|v_0|^+|a_0|^2}
\,,  \\
A_3 &=& -\frac{2\,{\rm Re}(v_1 a^*_1 + 2 v_0 a^*_0)}{|v_1|^2+|a_1|^2+|v_0|^+|a_0|^2}\,,
\end{eqnarray}
and two parity-even ratios
\begin{eqnarray}
\eta_1 &=& \frac{|v_1|^2+|a_1|^2-2|v_0|^2-2|a_0|^2}{
                 |v_1|^2+|a_1|^2+|v_0|^2+|a_0|^2}\,, \\
\eta_2 &=& \frac{2{\rm Re}(v_1v^*_0+a_1 a^*_0)}{
                 |v_1|^2+|a_1|^2+|v_0|^2+|a_0|^2}\,.
\end{eqnarray}
An explicit calculation leads to the following diagonal components of the
polar-angle distributions
\begin{eqnarray}
    W'_{\pm\pm}(\omega, \theta_Z)
&=& \frac{1}{2}\left[W'_T(\omega,\theta_Z) \pm Z'(\omega,\theta_Z)\right]\,,
\label{eq:z_polarization_density_matrix_1}\\
    W'_{\,0\,0}(\omega, \theta_Z)
&=& W'_L(\omega,\theta_Z)\,,
\label{eq:z_polarization_density_matrix_2}
\end{eqnarray}
with the parity-even and parity-odd transverse parts, $W'_T$ and $Z'_T$, and
the parity-even longitudinal part $W'_L$ given explicitly by
\begin{eqnarray}
  W'_T(\omega,\theta_Z)
&=&  \frac{2}{3} + \frac{1}{3} \cos\theta_Z P^T(3A_1-A_3)
   +\frac{1}{6}(3\cos^2\omega-1) \eta_1 \nonumber\\
&& +\frac{1}{6}(3\cos^2\omega-1)\cos\theta_Z P^T A_3
   +\frac{1}{\sqrt{2}}\cos\omega\sin\omega \sin\theta_Z P^T A_2\,, \\
  Z'_T(\omega, \theta_Z)
&=& \cos\omega A_1 +\frac{1}{3}\cos\omega \cos\theta_Z P^T(2+\eta_1)
   +\frac{1}{\sqrt{2}}\sin\omega\sin\theta_Z P^T\eta_2\,,  \\
  W'_L(\omega, \theta_Z)
&=& \frac{1}{3} + \frac{1}{6} \cos\theta_Z P^T(3A_1-A_3)
   -\frac{1}{6}(3\cos^2\omega-1) \eta_1 \nonumber\\
&& -\frac{1}{6}(3\cos^2\omega-1)\cos\theta_Z P^T A_3
   -\frac{1}{\sqrt{2}}\cos\omega\sin\omega \sin\theta_Z P^T A_2\,,
\end{eqnarray}
in the LAB. The diagonal elements are to be folded with the $Z$ decay
distributions in the $Z$ rest frame reconstructed directly from the LAB.\\

Among various decay modes of $Z$, the leptonic $Z$-boson decays
$Z\to\ell^-\ell^+$, in particular, with $\ell=e$ and $\mu$, provide us with a
very clean and powerful means for reconstructing the $Z$-boson rest frame,
independently of its production mechanism,  and extracting the information on
$Z$ polarization. The normalized $\ell^-$ polar-angle distributions with
respect to the $Z$ polarization defined to be the $Z$-boson momentum direction
in the LAB are given by
\begin{eqnarray}
  D_{\pm\pm}(\theta_\ell)
&=& \frac{1}{4}\left[1+\cos^2\theta_\ell \mp 2A_e \cos\theta_\ell\right]\,,
\label{eq:l_angular_dsitribution_1} \\
  D_{\,0\,0}(\theta_\ell)
&=& \frac{1}{2}(1-\cos^2\theta_\ell)\,.
\label{eq:l_angular_distribution_2}
\end{eqnarray}
Combining Eqs.$\,$(\ref{eq:z_polarization_density_matrix_1}) and
(\ref{eq:z_polarization_density_matrix_2}) with
Eqs.$\,$(\ref{eq:l_angular_dsitribution_1}) and
(\ref{eq:l_angular_distribution_2}) we can obtain the full spin and polar-angle
correlation of the two-stage decays $T\to Z t \to (\ell^-\ell^+) t$ as
\begin{eqnarray}
&&  \frac{d^2{\cal C}}{d\cos\theta_Z d\cos\theta_\ell}
  = \frac{1}{4}\bigg[ 1 +\frac{1}{2} P^T(3A_1-A_3) \cos\theta_Z
                     -\frac{3}{2} A_e A_1 \cos\omega\cos\theta_\ell
                     +\frac{1}{2}\eta_1P_2(\cos\omega)P_2(\cos\theta_\ell)
                     \nonumber\\
   &&\hskip 2.cm
      -\frac{1}{2}P^T A_e (2+\eta_1)\cos\omega\cos\theta_Z \cos\theta_\ell
      -\frac{3}{2\sqrt{2}}P^TA_e\eta_2\sin\omega\sin\theta_Z\cos\theta_\ell
      \nonumber\\
   &&\hskip 2.cm  +\frac{1}{2}P^TA_3P_2(\cos\omega)\cos\theta_Z
       P_2(\cos\theta_\ell)
      +\frac{3}{2\sqrt{2}}P^TA_2\cos\omega\sin\omega \sin\theta_Z
       P_2(\cos\theta_\ell)\bigg]\,,
\label{eq:theta-z_theta_l_correlation}
\end{eqnarray}
with the second Legendre polynomial $P_2(x)=(3x^2-1)/2$ introduced for
shortening the expression of the correlation function. As noted before, the
Wick helicity rotation angle $\omega$ is a function of $\theta_Z$, the polar
angle of $Z$ in the $T$ rest
frame, and two boost factors, $\beta_T$ and $\beta_Z$.\\

Folding the polar-angle correlation in
Eq.$\,$(\ref{eq:theta-z_theta_l_correlation}) with any given $T$ speed
distribution depending on a specific production mechanism yields the full
$(\theta_Z, \theta_\ell)$ correlation in the LAB. And integrating it over the
polar angle of the $Z$ boson we obtain the single
polar-angle distribution of the $\ell^-$ polar angle $\theta_\ell$.\\

\section{Conclusions}
\label{sec:conclusions}

In this work, we have provided a general and comprehensive spin analysis
through polar-angle correlations in any combinations of two-stage two-body
decays. {\it To summarize}, we have obtained the following key results from the
analysis:
\begin{itemize}
\item[$\ast$] A systematic review of the Wick helicity rotation on helicity
    states and decay helicity amplitudes was presented.
\item[$\ast$] Considering a two-body decay $X_2\to Y X_1$, we have described
    in detail how to transform through Wick helicity rotation the decay
    helicity amplitudes in the rest frame of the decaying particle $X_2$ to
    those in the LAB with the particle moving with non-zero speed.
\item[$\ast$] Combining the decay $X_2\to Y X_1$ and the sequential decay
    $Y\to ab$, we have derived the correlated distributions expressed in
    terms of the Wick helicity rotation angle, the $Y$ polar angle in the
    $X_2$RF and the $a$ polar angle in the $Y$RF. They can be applicable
    directly in the LAB.
\item[$\ast$] We have introduced polarization estimator functions with which
    all the observables depending on the $X_2$ polarization and the decay
    dynamical properties are conveniently expressed and so transparently
    described, even in the case when the $Y$ direction cannot be
    reconstructed event by event.
\item[$\ast$] For the sake of concrete demonstration, we have studied the
    characteristic tau-lepton pair polarization on the $Z$-boson pole and the
    top-quark pair production processes in $e^-e^+$ collisions in the
    framework of SM, and the decay of a heavy vectorlike top quark into a top
    and a $Z$-boson expected to occur in some models beyond the SM such as
    little Higgs models.
\item[$\ast$] For completeness, all the useful formulas directly applicable
    for any spin and polar-angle correlations in any two-stage two-body
    decays are collected and explained in some detail.
\end{itemize}

Generally, a (new) heavy particle decays in a series of stages, often,
including two-stage two-body decays. In this situation, the formalism presented
in the present work will be very useful and powerful in determining all the
particle spins in the processes and probing their dynamical properties. Based
on the formalism, more interesting and concrete examples will be studied and
presented in a forthcoming work.\\


\subsection*{Acknowledgments}

The work was supported in part by Basic Science Research Program through the
National Research Foundation (NRF) funded by the Ministry of Education, Science
and Technology (NRF-2016R1D1A3B01010529) program and in part by the CERN-Korea
theory collaboration.

\vskip 1.0cm

\appendix

\section{Wigner \boldmath{$D$}- and \boldmath{$d$}-functions}
\label{appendix:wigner_d_functions}

\setcounter{equation}{0}
\renewcommand{\theequation}{\thesection.\arabic{equation}}

Let $J_x, J_y, J_z$ be three angular momentum generators in a fixed rectangular
coordinate system. The Casimir operator $J^2= J^2_x+J^2_y+J^2_z$ commutes with
all angular momentum generators and it can be diagonalized together with $J_z$,
forming a complete set of orthogonal eigenstates with
\begin{eqnarray}
J^2\,|j\lambda\rangle = j(j+1)\,|j\lambda\rangle
\quad \mbox{and} \quad
J_z\,|j\lambda\rangle = \lambda\, |j\lambda\rangle\,,
\end{eqnarray}
where $j=0. 1/2, 1, 3/2, \cdots$ and $\lambda = -j, -j+1,\cdots, j$ for a given
$j$. The angular momentum operators can be used to define a three-dimensional rotation
operator ${\cal R}(\alpha, \beta, \gamma)$ as
\begin{eqnarray}
  {\cal R}(\alpha,\beta,\gamma)
= e^{-i\alpha J_z} e^{-i\beta J_y} e^{-i\gamma J_z}\,,
\label{eq:3-dim_rotation_operator}
\end{eqnarray}
where $\alpha, \beta, \gamma$ are Euler angles (characterized by the right-handed and
active interpretation). \\

The Wigner $D$-functions are the matrix elements of the rotation operator
$\mathcal{R}$ in Eq.$\,$(\ref{eq:3-dim_rotation_operator}) of which the
explicit form is
\begin{eqnarray}
  D^j_{\sigma,\lambda} (\alpha,\beta,\gamma)
= \langle j\sigma| {\cal R}(\alpha, \beta, \gamma)|j\lambda\rangle
= e^{-i\sigma \alpha}\, d^j_{\sigma,\lambda}(\beta)\, e^{-i\lambda\gamma}\,,
\end{eqnarray}
where the mutually orthogonal Wigner $d$-functions are the matrix elements
defined as
\begin{eqnarray}
d^j_{\sigma,\lambda}(\beta) = \langle j\sigma| e^{-i\beta J_y}|j\lambda\rangle\,,
\end{eqnarray}
which are real. By definition the orthogonal $d$-functions satisfy the group
properties:
\begin{eqnarray}
    \sum_{\lambda} d^j_{\mu,\lambda}(\beta)\, d^j_{\nu,\lambda}(\beta)
&=& \sum_{\sigma} d^j_{\sigma,\mu}(\beta)\, d^j_{\sigma,\nu}(\beta)
\, =\,  \delta_{\mu\nu}\,,  \\
    d^j_{\sigma,\lambda} (\beta_1+\beta_2)
&=& \sum_{\mu} d^j_{\sigma,\mu}(\beta_1)\,\,
               d^j_{\mu,\lambda}(\beta_2)\,,
\end{eqnarray}
reflecting the characteristic additive property of two successive rotations.\\

For the sake of convenient discussion, the expressions of Wigner $d$-functions
for the spin-1/2 and spin-1 cases are listed explicitly in
Tab.~\ref{tab:explicit_d_functions}.\\

\begin{table}[htb]
\centering
\renewcommand{\arraystretch}{1.35}
\begin{tabular}{|c|c|c|}
\hline
\multicolumn{3}{|c| }{\boldmath{$d^{1/2}_{\sigma,\lambda}(\theta)$}} \\
\hline\hline
{\small\backslashbox{$\sigma$}{$\lambda$}}
               & $\frac{1}{2}$
               & $-\frac{1}{2}$ \\ \hline
$\phantom{+}\frac{1}{2}$  & $\cos\frac{\theta}{2}$
               & $-\sin\frac{\theta}{2}$ \\
$-\frac{1}{2}$ & $\sin\frac{\theta}{2}$
               & $\phantom{+}\cos\frac{\theta}{2}$  \\
\hline
\end{tabular}
\hskip 1.5cm
\begin{tabular}{|c|c|c|c|}
\hline
\multicolumn{4}{|c|}{\boldmath{$d^1_{\sigma,\lambda}(\theta)$}} \\
\hline\hline
{\small\backslashbox{$\sigma$}{$\lambda$}}
               & $1$
               & $0$
               & $-1$ \\ \hline
$\phantom{+} 1$  & $\frac{1+\cos\theta}{2}$
                 & $-\frac{\sin\theta}{\sqrt{2}}$
                 & $\frac{1-\cos\theta}{2}$ \\
$\phantom{+} 0$  & $\frac{\sin\theta}{\sqrt{2}}$
                 & $\cos\theta$
                 & $-\frac{\sin\theta}{\sqrt{2}}$ \\
$-1$  & $\frac{1-\cos\theta}{2}$
                 & $\frac{\sin\theta}{\sqrt{2}}$
                 & $\frac{1+\cos\theta}{2}$ \\
\hline
\end{tabular}
\vskip 0.5cm
\caption{\it Expressions of $d$ functions in terms of a polar angle $\theta$
          for two spin values,  $j=1/2$ (left table) and $j=1$ (right table),
          which are used in the present work.
          }
\label{tab:explicit_d_functions}
\end{table}

The Wigner $D$-functions form a set of orthogonal functions of the Euler angles:
\begin{eqnarray}
  \int^{2\pi}_{0} d\alpha \int^{1}_{-1} d\cos\beta \int^{2\pi}_{0} d\gamma\,\,
  D^{j'*}_{\sigma',\lambda'} (\alpha, \beta,\gamma)\,
  D^j_{\sigma,\lambda} (\alpha,\beta,\gamma)
= \frac{8\pi^2}{2j+1}\, \delta_{j'j}\, \delta_{\sigma'\sigma}\, \delta_{\lambda'\lambda}
\,,
\end{eqnarray}
leading to the orthogonal condition for the $d$-functions
\begin{eqnarray}
  \int^{1}_{-1} d\cos\beta\,\,
  d^{j'}_{\sigma,\lambda} (\beta)\,\,
  d^j_{\sigma,\lambda} (\beta)
= \frac{2}{2j+1}\, \delta_{j'j}\,.
\end{eqnarray}
In addition, the $d$-functions enjoy several symmetry properties:
\begin{eqnarray}
  d^j_{\sigma,\lambda}(\beta)
&=& (-1)^{\sigma-\lambda}\, d^j_{\lambda,\sigma}(\beta)\,,\\
  d^j_{\sigma,\lambda}(\beta)
&=& d^j_{\lambda,\sigma}(-\beta)\,, \\
  d^j_{\sigma,\lambda}(\beta)
&=& d^j_{-\lambda,-\sigma}(\beta)\,, \\
  d^j_{\sigma,\lambda}(\pi-\beta)
&=& (-1)^{j+\sigma}\, d^j_{\sigma,-\lambda} (\beta)\,,
\end{eqnarray}
and they satisfy two useful coupling rules involving Clebsch-Gordan coefficients:
\begin{eqnarray}
&& d^{j_1}_{\sigma_1,\lambda_1}(\beta)\,
  d^{j_2}_{\sigma_2,\lambda_2}(\beta)
= \sum_{j=|j_1-j_2|}^{j_1+j_2}
  \langle j_1\sigma_1 j_2\sigma_2 |j \sigma_+\rangle\,
  \langle j_1\lambda_1 j_2\lambda_2 | j\lambda_+\rangle\,
  d^{j}_{\sigma_+,\lambda_+}(\beta)\,,
\label{eq:d_coupling_rule_1}\\
&& d^{j_1}_{\sigma_1,\lambda_1}(\beta)\,
  d^{j_2}_{\sigma_2,\lambda_2}(\beta)
= \sum_{j=|j_1-j_2|}^{j_1+j_2}
  \langle j_1\sigma_1 j_2,-\sigma_2 |j \sigma_-\rangle\,
  \langle j_1\lambda_1 j_2,-\lambda_2 | j\lambda_-\rangle\,\nonumber\\
&& \mbox{ }\hskip 5.cm \times (-1)^{\sigma_2-\lambda_2}\,
    d^{j}_{\sigma_-,\lambda_-}(\beta)\,,
\label{eq:d_coupling_rule_2}
\end{eqnarray}
with the constraints $\sigma_\pm=\sigma_1\pm \sigma_2$ and
$\lambda_\pm=\lambda_1\pm \lambda_2$. The bracket expression
$\langle j_1\mu_1 j_2\mu_2 |j \mu\rangle$ is a Clebsch-Gordan
coefficient. Two useful properties of the Clebsch-Gordan coefficients
are
\begin{eqnarray}
  \sum_{\sigma}(-1)^\sigma \langle j \sigma j (-\sigma)|J0\rangle
=  (-1)^j \sqrt{2j+1}\,\, \delta_{J0} \ \ \mbox{with}\ \
  \langle j\sigma j (-\sigma)|00\rangle
= \frac{(-1)^{j-\sigma}}{\sqrt{2j+1}}\,,
\end{eqnarray}
with which the orthogonality relations of $d$-functions can be easily derived.

\section{Wick helicity rotation distribution functions (WDFs)}
\label{appendix:wick_rotation_distribution_functions}

\setcounter{equation}{0}
\renewcommand{\theequation}{\thesection.\arabic{equation}}

Before exhibiting a set of WDFs defined in
Eq.$\,$(\ref{eq:wick_distribution_functions}) in their explicit form for the
spin values up to $1$ in this appendix, we emphasize that the formalism given
in the main text is so general that it can be applied to any combination of the
spins, $\{j_2, j, j_1\}$, of the particles, $\{X_2, Y, X_1\}$, in the decay
$X_2\to Y X_1$ in a model-independent manner. Instead of any detailed
derivations, which are demonstrated with a few examples in the main text, the
essential parts for deriving WDFs and the resulting $Y$ polarization density
matrices are
collected in this Appendix.\\

\subsection{\fbox{\boldmath{$j_2 \to 0 + j_1$}}}
\label{subsec:j2_0_j1}

The simplest case is for the $Y$ particle of zero spin ($j=0$), because of no
Wick helicity rotation effects at all. The (unnormalized) WDF simply reads
\begin{eqnarray}
  {\cal D}'(\omega,\theta)
= \sum_{\sigma_2} \sum_{\sigma_1}
  \rho^{X_2}_{\sigma_2,\sigma_2}
  [d^{j_2}_{\sigma_2,-\sigma_1}(\theta)]^2\,
  |F^{j_2}_{\sigma_1}|^2
  \ \ \Rightarrow \ \
  \langle {\cal D}'(\omega,\theta)\rangle
= \frac{1}{(2j_2+1)}
   {\sum_{\sigma_1}}' |F^{j_2}_{\sigma_1}|^2\,,
\label{eq:j2_0_j1_correlated_distribution}
\end{eqnarray}
with the sum over $\sigma_1$ satisfying the constraint $|\sigma_1|\leq j_2$.\\

The hadronic decay processes $\Sigma \to \pi^- p$ and $\rho^-\to\pi^-\pi^0$ and
any two-body decay involving a Higgs boson $H$ belong to this category of
two-body decays.\\

\subsection{\fbox{\boldmath{$0 \to 1/2 + 1/2$}}}
\label{subsec:0_1/2_1/2}

The first non-trivial Wick helicity rotation effects show up in the case with
$j_2=0$ and $j=j_1=1/2$. Two typical examples of this category are $H\to
\tau^-\tau^+$ in the SM and $H^-\to \tau^- \bar{\nu}_\tau$ in a two-Higgs
doublet model
\cite{Lee:1973iz,Branco:2011iw}.\\

An explicit calculation of the WDFs in this case leads to the expression:
\begin{eqnarray}
  {\cal D}'_{\sigma'\lambda'}(\omega, \theta)
= \sum_{\sigma=-1/2}^{1/2}
  [d^{1/2}_{\sigma',\sigma}(\omega)\,
   d^{1/2}_{\lambda',\sigma}(\omega)]\,\,
  |F^0_{\sigma\sigma}|^2
  \ \ \Rightarrow \ \
  {\rm Tr}({\cal D}') = \sum_{\sigma} |F^0_{\sigma\sigma}|^2\,,
\end{eqnarray}
leading to the angle-dependent $2\times 2$ distributions, density matrix
$W'(\omega,\theta)$ of the particle $Y$ in the LAB, as
\begin{eqnarray}
  W'(\omega, \theta)
= \frac{1}{2}
  \left(\begin{array}{cc}
        1+ \cos\omega\, A & \sin\omega\, A \\[2mm]
        \sin\omega\, A & 1-\cos\omega\, A
        \end{array}\right)
\quad \mbox{with} \quad
 A = \frac{|F^0_{++}|^2-|F^0_{--}|^2}{
             |F^0_{++}|^2-|F^0_{--}|^2}\,,
\end{eqnarray}
in the $(1/2, -1/2)$ helicity basis, where $A$ is the $Y$ polarization in the
$X_2$RF given in terms of the reduced helicity elements  $F^0_{\pm \frac{1}{2}
\pm \frac{1}{2}}$ denoted by the simplified notations $F^0_{\pm\pm}$. The
diagonal elements of the $Y$ angular distribution averaged over the polar angle
$\theta$ are given by the polarization estimator $\langle \cos\omega \rangle$,
of which the expressions are given in terms of $\beta_2$ and $\beta$ in
Appendix~\ref{appendix:polarization_estimator_functions} and the parity-odd
ratio $A$ as
\begin{eqnarray}
  \rho^Y_{\pm\pm}
= \frac{1}{2} \bigg[ 1 \pm \langle \cos\omega \rangle\, A\bigg]\,.
\end{eqnarray}
For example, the decay process $H\to \tau^-\tau^+$ with a possible
parity-violating $H\tau\tau$ coupling and a fixed $H$ energy as in the
Higgsstrahlung process $e^+e^-\to
HZ$ is an interesting example of this decay category. \\

\subsection{\fbox{\boldmath{$0 \to 1 + 0$}}}
\label{subsec:0_1_0}

One interesting example of this type of decays is $\tilde{t}_2 \to Z \tilde{t}_1$
in the minimal supersymmetric Standard Model (MSSM), which can be realized if the mass
difference between two top squarks is larger than the $Z$-boson mass $m_Z$.
We note that rotational invariance forces the $Z$ boson to be longitudinally
polarized.\\

An explicit calculation of the WDFs in this spin-1 $Y$ case leads to the expression:
\begin{eqnarray}
  {\cal D}'_{\sigma'\lambda'}(\omega, \theta)
= [d^{1}_{\sigma',0}(\omega)\,
   d^{1}_{\lambda',0}(\omega)]\,\,
  |F^0_0|^2
  \ \ \Rightarrow \ \
  {\rm Tr}({\cal D}') = \sum_{\sigma} |F^0_0|^2\,,
\end{eqnarray}
leading to the angle-dependent $3\times 3$ distributions, $W'$ of the particle
$Y$ in the LAB as
\begin{eqnarray}
  W'(\omega, \theta)
= \frac{1}{2}
  \left(\begin{array}{ccc}
        \sin^2\omega  & -\sqrt{2}\sin\omega\cos\omega & -\sin^2\omega \\[1mm]
        -\sqrt{2}\sin\omega\cos\omega & 2 \cos^2\omega & \sqrt{2}\sin\omega\cos\omega
         \\[1mm]
        -\sin^2\omega & \sqrt{2}\sin\omega\cos\omega & \sin^2\omega
        \end{array}\right)\,,
\end{eqnarray}
in the $(1,0,-1)$ basis, independent of any dynamical parameters involved in
the decay and also with no explicit $\theta$-dependence. The diagonal elements
of the $Y$ angular distribution averaged over the polar angle $\theta$ are
given by the PEF $\langle \cos^2\omega \rangle$, of which the explicit form is
given in Appendix~\ref{appendix:polarization_estimator_functions}, as
\begin{eqnarray}
  \rho^Y_{\pm\pm}
= \frac{1}{3}-\frac{1}{6}\langle\, 3 \cos^2\omega-1\,\rangle
\quad\mbox{and}\quad
  \rho^Y_{00} = \frac{1}{3} + \frac{1}{3} \langle\, 3\cos^2\omega-1\,\rangle\,.
\end{eqnarray}
An example of this category is the decay process $\tilde{f}_2\to Z \tilde{f}_1$
assuming that $\tilde{f}_2$ is produced in association with $\tilde{f}_1^*$
through the process $e^+e^-\to\tilde{f}_2\, \tilde{f}^*_1$ of any flavor of
sfermions, which may be realized in the MSSM.\\

\subsection{\fbox{\boldmath{$0 \to 1 + 1$}}}
\label{subsec:0_1_1}

Although it is a loop-induced process and so its branching ratio is small,
one important process of this decay type is the radiative decay $H\to Z \gamma$
in the SM and its extensions.\\

An explicit calculation of the WDFs in this case gives us the expression for
the WDFs
\begin{eqnarray}
  {\cal D}'_{\sigma'\lambda'}(\omega, \theta)
= \sum_{\sigma=-1}^{1}\,
  [\, d^{1}_{\sigma',\sigma}(\omega)\,
   d^{1}_{\lambda',\sigma}(\omega)]\,\,
  |F^0_{\sigma\sigma}|^2
  \ \ \Rightarrow \ \
  {\rm Tr}({\cal D}') = \sum_{\sigma} |F^0_{\sigma\sigma}|^2\,.
\end{eqnarray}
For the sake of notation, we introduce a parity-odd polarization parameter
$P_R$ and a parity-even polarization parameter as
\begin{eqnarray}
A &=& \frac{|F^0_{++}|^2-|F^0_{--}|^2}{|F^0_{++}|^2+|F^0_{00}|^2+|F^0_{--}|^2}\,,
  \\[2mm]
\eta &=& \frac{|F^0_{++}|^2-2|F^0_{00}|^2+|F^0_{--}|^2}{
              |F^0_{++}|^2+|F^0_{00}|^2+|F^0_{--}|^2}\,.
\end{eqnarray}
Note that $\eta=1$ if the $X_1$ particle is a photon $\gamma$ with no
longitudinal mode. Three diagonal elements of the $3\times 3$ density matrix
$\rho^Y$ averaged over the polar angle $\theta$ are given in terms of the
parameters by
\begin{eqnarray}
  \rho^Y_{\pm\pm}
&=& \frac{1}{3}\pm \frac{2}{3} \langle \cos\omega \rangle\, A
                    + \frac{1}{12} \langle\, 3\cos^2\omega-1\,\rangle\, \eta \\
&&                    \ \ \Rightarrow \ \
                    \frac{1}{3}+\frac{1}{12}\langle 3\cos^2\omega-1\rangle)
                    \pm \frac{2}{3} \langle \cos\omega \rangle\, A
                    \ \ \mbox{for} \ \ \eta=1\,,\\
 \rho^Y_{00}
&=& \frac{1}{3}-\frac{1}{6}\langle\, 3\cos^2\omega-1\,\rangle\, \eta \\
&&                    \ \ \Rightarrow \ \
                    \frac{1}{3}-\frac{1}{6}\langle\, 3\cos^2\omega-1\,\rangle)
                    \ \ \mbox{for} \ \ \eta=1\,.
\end{eqnarray}
Furthermore for the two-photon modes such as $H\to\gamma\gamma$ and
$\pi^0\to \gamma\gamma$, the longitudinal diagonal element $\rho^Y_{00}$
cannot exist as indicated by $\eta=1$ as well as $\omega=0$ for massless particles.\\

\subsection{\fbox{\boldmath{$1/2 \to 1/2 + 0$}}}
\label{subsec:1/2_1_1/2}

This category contains the hyperon decays, $\Lambda \to p \pi^-$ and
$\Lambda\to n \pi^0$, in the SM and the decay of the second
lightest neutralino, $\tilde{\chi}^0_2\to t \tilde{t}^*_1$ in the MSSM,
if kinematically allowed.\\

The helicity amplitude of this decay mode in the rest frame of $X_2$
is of the form
\begin{eqnarray}
  {\cal D}_{\sigma_2\sigma}
= F^{1/2}_\sigma\, d^{1/2}_{\sigma_2\sigma}(\theta)\,,
\end{eqnarray}
with the $X_2$ and $Y$ helicities, $\sigma_2=\pm 1/2=\pm$ and $\sigma=\pm 1/2=\pm$,
and the WDFs in the LAB, where the parent particle $X_2$ move with speed $\beta_2$,
read
\begin{eqnarray}
  {\cal D}'_{\sigma'\lambda'}(\omega,\theta)
= \sum_{\sigma'_2} \sum_{\sigma,\lambda}\,\,
  [\, d^{1/2}_{\sigma'\sigma}(\omega) d^{1/2}_{\lambda'\lambda}(\omega)]\,
  [d^{1/2}_{\sigma'_2\sigma}(\theta) d^{1/2}_{\sigma'_2\lambda}(\theta)]\,
  F^{1/2}_{\sigma}F^{1/2*}_{\lambda}\,.
\end{eqnarray}
For notational convenience, we introduce a parity-odd polarization parameter
$A$ and a parity-even polarization parameter $\eta$ as
\begin{eqnarray}
A &=& \frac{|F^{1/2}_+|^2-|F^{1/2}_-|^2}{
            |F^{1/2}_+|^2+|F^{1/2}_-|^2}\,, \\
\eta &=& \frac{2{\rm Re}(F^{1/2}_+ F^{1/2*}_-)}{
            |F^{1/2}_+|^2+|F^{1/2}_-|^2}\,.
\end{eqnarray}
In terms of the parameters $A$ and $\eta$ can we derive two diagonal elements
and thus the degree of longitudinal polarization $P_L$ in the LAB as
\begin{eqnarray}
  \rho^Y_{\pm\pm}
&=& \frac{1}{2}
   \bigg[1\pm \left(\langle \cos\omega\rangle A
               +\langle \cos\omega \cos\theta\rangle P^{X_2}
               +\langle \sin\omega \sin\theta\rangle P^{X_2}\eta\right)
               \bigg]\,,\\
   P_L
&=& \langle \cos\omega\rangle A
               +\langle \cos\omega \cos\theta\rangle P^{X_2}
               +\langle \sin\omega \sin\theta\rangle P^{X_2}\eta\,.
\end{eqnarray}
Another interesting example of this category is the decay $T\to tH$ of a new
heavy top quark $T$ into a top quark $t$ and a Higgs boson $H$ in
the little Higgs models.\\

\subsection{\fbox{\boldmath{$1/2 \to 1/2 + 1$}}}
\label{subsec:1/2_1/2_1}

An interesting example of this category is the decay $T\to t Z$ of a new heavy
top quark $T$ into a $Z$ boson and a top quark $t$ in the littlest Higgs model,
one of the popular models beyond the SM. \\

The helicity amplitude of this decay mode in the $X_2$RF is of the form
\begin{eqnarray}
  D_{\sigma_2;\sigma,\sigma_1}
= F^{1/2}_{\sigma\sigma_1}\, d^{1/2}_{\sigma_2,\sigma-\sigma_1}(\theta)\,,
\end{eqnarray}
with the $X_2$, $X_1$ and $Y$ helicities, $\sigma_2=\pm 1/2=\pm$, $\sigma_1=\pm 1, 0$,
and $\sigma=\pm 1/2=\pm$, and the WDFs in the LAB, where the parent particle $X_2$
move with speed $\beta_2$, read
\begin{eqnarray}
  {\cal D}'_{\sigma'\lambda'}(\omega,\theta)
= \sum_{\sigma'_2} \sum_{\sigma,\lambda} \sum_{\sigma_1}\,
  [\, d^{1/2}_{\sigma'\sigma}(\omega) d^{1/2}_{\lambda'\lambda}(\omega)]\,
  [d^{1/2}_{\sigma'_2,\sigma-\sigma_1}(\theta)\,
   d^{1/2}_{\sigma'_2,\lambda-\sigma_1}(\theta)]\,
  F^{1/2}_{\sigma\sigma_1}F^{1/2*}_{\lambda\sigma_1}\,.
\end{eqnarray}
For notational convenience, we introduce a parity-odd polarization parameter
$P_R$ and two parity-even polarization parameters, $\eta_{1R}$ and $\eta_{2R}$,
as
\begin{eqnarray}
A &=& \frac{|F^{1/2}_{++}|^2+|F^{1/2}_{+0}|^2
             -|F^{1/2}_{--}|^2-|F^{1/2}_{-0}|^2}{
              |F^{1/2}_{++}|^2+|F^{1/2}_{+0}|^2
             +|F^{1/2}_{--}|^2+|F^{1/2}_{-0}|^2}\,,\\
\eta_{1} &=& \frac{|F^{1/2}_{++}|^2-|F^{1/2}_{+0}|^2
             +|F^{1/2}_{--}|^2-|F^{1/2}_{-0}|^2}{
              |F^{1/2}_{++}|^2+|F^{1/2}_{+0}|^2
             +|F^{1/2}_{--}|^2+|F^{1/2}_{-0}|^2}\,,\\
\eta_{2} &=& \frac{2{\rm Re}(F^{1/2}_{+0} F^{1/2*}_{-0})}{
              |F^{1/2}_{++}|^2+|F^{1/2}_{+0}|^2
             +|F^{1/2}_{--}|^2+|F^{1/2}_{-0}|^2}\,.
\end{eqnarray}
In terms of the polarization parameters, $A$ and $\eta_{1, 2}$ we can derive
two diagonal elements and thus the degree of longitudinal polarization $P_L$ in
the LAB as
\begin{eqnarray}
  \rho^Y_{\pm\pm}
&=& \frac{1}{2}
   \bigg[1\pm \left(\langle \cos\omega\rangle A
               -\langle \cos\omega \cos\theta\rangle P^{X_2}\eta_{1}
               +\langle \sin\omega \sin\theta\rangle P^{X_2}\eta_{2}\right)
               \bigg]\,,\\
   P_L
&=& \langle \cos\omega\rangle A
               -\langle \cos\omega \cos\theta\rangle P^{X_2}\eta_{1}
               +\langle \sin\omega \sin\theta\rangle P^{X_2}\eta_{2}\,.
\end{eqnarray}
The two-body decay $T\to tZ$ of a new heavy top quark $T$ into a top
quark $t$ and a $Z$ in the little Higgs model is studied in detail
as a characteristic example of this category in
Subsection~\ref{subsec:T_Zt_llt}.\\

\subsection{\fbox{\boldmath{$1/2 \to 1 + 1/2$}}}
\label{subsec:1/2_1_1/2}

This category contains several SM examples such as $t\to W^+ b$, $\tau^-\to
\rho^-\nu_\tau$, and $\tau^-\to a^-\nu_\tau$ as well as the loop-induced
flavor-changing processes such as $t\to Z c$.\\

In the $X_2$ rest frame, the helicity amplitude can be cast into the
form:
\begin{eqnarray}
  D_{\sigma_2;\sigma,\sigma_1}
= F^{1/2}_{\sigma\sigma_1}\,\, d^{1/2}_{\sigma_2, \sigma-\sigma_1}(\theta)\,
  e^{i\sigma_2\phi}\,,
\end{eqnarray}
where the $X_2$ helicity $\sigma_2=\pm 1/2=\pm$, the $Y$ helicity
$\sigma=\pm 1, 0 =\pm, 0$ and the $X_1$ helicity $\sigma_1=\pm 1/2 =\pm$.
Note that the amplitudes $F^{1/2}_{+-}$ and $F^{1/2}_{-+}$ are forbidden
due to angular momentum conservation.\\

For notational convenience, let us introduce three parity-odd polarization
parameters defined as
\begin{eqnarray}
    A_{1}
&=& \frac{|F^{1/2}_{++}|^2-|F^{1/2}_{--}|^2}{
    |F^{1/2}_{++}|^2+|F^{1/2}_{--}|^2+|F^{1/2}_{0+}|^2+|F^{1/2}_{0-}|^2}\,,
    \\
    A_{2}
&=& \frac{2{\rm Re}(F^{1/2}_{++} F^{1/2*}_{0+} - F^{1/2}_{--} F^{1/2*}_{0-})}{
    |F^{1/2}_{++}|^2+|F^{1/2}_{--}|^2+|F^{1/2}_{0+}|^2+|F^{1/2}_{0-}|^2}\,,
    \\
    A_{3}
&=& \frac{|F^{1/2}_{++}|^2-|F^{1/2}_{--}|^2+2|F^{1/2}_{0+}|^2-2|F^{1/2}_{0-}|^2}{
    |F^{1/2}_{++}|^2+|F^{1/2}_{--}|^2+|F^{1/2}_{0+}|^2+|F^{1/2}_{0-}|^2}\,,
\end{eqnarray}
and two parity-even polarization parameters
\begin{eqnarray}
    \eta_{1}
&=& \frac{|F^{1/2}_{++}|^2+|F^{1/2}_{--}|^2-2|F^{1/2}_{0+}|^2-2|F^{1/2}_{0-}|^2}{
    |F^{1/2}_{++}|^2+|F^{1/2}_{--}|^2+|F^{1/2}_{0+}|^2+|F^{1/2}_{0-}|^2}\,,
    \\
    \eta_{2}
&=& \frac{2{\rm Re}(F^{1/2}_{++} F^{1/2*}_{0+} + F^{1/2}_{--} F^{1/2*}_{0-})}{
    |F^{1/2}_{++}|^2+|F^{1/2}_{--}|^2+|F^{1/2}_{0+}|^2+|F^{1/2}_{0-}|^2}\,.
\end{eqnarray}
Three diagonal elements of the $3\times 3$ density matrix $\rho^Y$ averaged
over the polar-angle distribution $d\Gamma/d\theta$ are given in terms of
the five polarization parameters by
\begin{eqnarray}
  \rho^Y_{\pm\pm}
&=& \frac{1}{3}\pm \frac{1}{2}\langle \cos\omega\rangle A_{1}
   +\frac{1}{12} \langle 3\cos^2\omega-1\rangle \eta_{1} \nonumber\\
&& +\frac{1}{2\sqrt{2}} \langle \cos\omega \sin\omega \sin\theta\rangle
    P^{X_2} A_{2}
   +\frac{1}{12} \langle (3\cos^2\omega-1)\cos\theta\rangle
    P^{X_2} A_{3} \nonumber\\
&& \pm \frac{1}{6}\langle \cos\omega\cos\theta\rangle P^{X_2}(2+\eta_{1})
   \pm \frac{1}{2\sqrt{2}} \langle \sin\omega\sin\theta \rangle P^{X_2} \eta_{2}\,, \\
  \rho^Y_{00}
&=& \frac{1}{3}-\frac{1}{6} \langle 3\cos^2\omega-1\rangle \eta_{1} \nonumber\\
&& -\frac{1}{\sqrt{2}} \langle \cos\omega \sin\omega \sin\theta\rangle
    P^{X_2} A_{2}
   -\frac{1}{6} \langle (3\cos^2\omega-1)\cos\theta\rangle
    P^{X_2} A_{3}\,,
\end{eqnarray}
satisfying the normalization condition
${\rm Tr}(\rho^Y)= \rho^Y_{++}+\rho^Y_{00}+\rho^Y_{--}=1$.\\

\subsection{\fbox{\boldmath{$1 \to 1/2 + 1/2$}}}
\label{subsec:1_1/2_1/2}

This decay category contains the SM processes such as the parity-violating
weak decays of the massive weak bosons, $W^+\to \tau^+ \nu_\tau$ and $Z\to\tau^-\tau^+$.\\

In the $X_2$ rest frame, the helicity amplitude of this type of decay modes
can be cast into the form:
\begin{eqnarray}
  D_{\sigma_2;\sigma,\sigma_1}
= F^{1}_{\sigma\sigma_1} d^1_{\sigma_2, \sigma-\sigma_1}(\theta)
  e^{i\sigma_2\phi}\,,
\end{eqnarray}
where the $X_2$ helicity $\sigma_2=\pm 1, 0=\pm, 0$, the $Y$ helicity
$\sigma=\pm 1/2=\pm$ and the $X_1$ helicity $\sigma_1=\pm 1/2 =\pm$.\\

For notational convenience, let us introduce three parity-odd polarization
parameters defined as
\begin{eqnarray}
    A_{1}
&=& \frac{|F^1_{++}|^2-|F^1_{--}|^2+|F^1_{+-}|^2-|F^1_{-+}|^2}{
    |F^1_{++}|^2+|F^1_{--}|^2+|F^1_{+-}|^2+|F^1_{-+}|^2}\,,
    \\
    A_{2}
&=& \frac{2{\rm Re}(F^1_{++} F^{1*}_{-+} - F^1_{--} F^{1*}_{+-})}{
    |F^1_{++}|^2+|F^1_{--}|^2+|F^1_{+-}|^2+|F^1_{-+}|^2}\,,
    \\
    A_{3}
&=& \frac{|F^1_{+-}|^2-|F^1_{-+}|^2-2|F^1_{++}|^2+2|F^1_{--}|^2}{
    |F^1_{++}|^2+|F^1_{--}|^2+|F^1_{+-}|^2+|F^1_{-+}|^2}\,,
\end{eqnarray}
and two parity-even polarization parameters
\begin{eqnarray}
    \eta_{1}
&=& \frac{|F^1_{+-}|^2+|F^1_{-+}|^2}{
    |F^1_{++}|^2+|F^1_{--}|^2+|F^1_{+-}|^2+|F^1_{-+}|^2}\,,
    \\
    \eta_{2}
&=& \frac{2{\rm Re}(F^1_{++} F^{1*}_{-+} + F^1_{--} F^{1*}_{+-})}{
    |F^1_{++}|^2+|F^1_{--}|^2+|F^1_{+-}|^2+|F^1_{-+}|^2}\,.
\end{eqnarray}
Two diagonal elements of the $2\times 2$ density matrix $\rho^Y$ averaged
over the polar-angle distribution $d\Gamma/d\theta$ are given in terms of
the five polarization parameters by
\begin{eqnarray}
  \rho^Y_{\pm\pm}
&=& \frac{1}{2}\pm \frac{1}{2}\langle \cos\omega\rangle A_{1}
   \pm \frac{1}{8} \langle \cos\omega(3\cos^2\theta-1)\rangle
    Q^{X_2} A_{3} \nonumber\\
&& \pm \frac{3}{4} \langle \cos\omega \cos\theta \rangle
    P^{X_2} \eta_{1}
   \pm \frac{3}{4\sqrt{2}} \langle\sin\omega\sin\theta\rangle
    P^{X_2} \eta_{2} \nonumber\\
&& \mp \frac{3}{4\sqrt{2}}\langle \sin\omega\cos\theta\sin\theta\rangle\,
   Q^{X_2} A_{2}\,,
\end{eqnarray}
with the longitudinal polarization $P^{X_2}=\rho^{X_2}_{++}-\rho^{X_2}_{--}$
and the (diagonal) tensor polarization $Q^{X_2}=\rho^{X_2}_{++}
+\rho^{X_2}_{--}-2\rho^{X_2}_{00}$ of the decaying particle $X_2$,
satisfying the normalization condition ${\rm Tr}(\rho^Y)=
\rho^Y_{++}+\rho^Y_{--}=1$.\\

\subsection{\fbox{\boldmath{$1 \to 1 + 0$}}}
\label{subsec:1_1_0}

The process $Z\to W^\pm\pi^\mp$ in the SM might be an interesting example of
this decay category, which is yet to be confirmed experimentally. A
non-standard example is the decay of a heavy vector boson $W^\pm_H$ into a SM
gauge boson and a SM Higgs boson such as $W^\pm_H \to W^\pm H$, appearing
in the little Higgs models \cite{Han:2003wu}. \\

In the $X_2$ rest frame, the helicity amplitude of this type of decay modes
can be cast into the form:
\begin{eqnarray}
  D_{\sigma_2;\sigma}
= F^{1}_{\sigma\sigma_1} d^1_{\sigma_2, \sigma}(\theta)\,
  e^{i\sigma_2\phi}\,,
\end{eqnarray}
where the $X_2$ helicity $\sigma_2=\pm 1, 0=\pm, 0$, the $Y$ helicity
$\sigma=\pm 1, 0=\pm, 0$ while the $X_1$ particle is spinless.\\

For notational convenience, let us introduce two parity-odd polarization
parameters defined as
\begin{eqnarray}
    A_{1}
&=& \frac{|F^1_{+}|^2-|F^1_{-}|^2}{
    |F^1_{+}|^2+|F^1_{0}|^2+|F^1_{-}|^2}\,,
    \\
    A_{2}
&=& \frac{2{\rm Re}(F^1_{+} F^{1*}_{0} - F^1_{-} F^{1*}_{0})}{
    |F^1_{+}|^2+|F^1_{0}|^2+|F^1_{-}|^2}\,,
\end{eqnarray}
and three parity-even polarization parameters
\begin{eqnarray}
    \eta_{1}
&=& \frac{|F^1_{+}|^2-2|F^1_{0}|^2+|F^1_{-}|^2}{
    |F^1_{+}|^2+|F^1_{0}|^2+|F^1_{-}|^2}\,,
    \\
    \eta_{2}
&=& \frac{2{\rm Re}(F^1_{+} F^{1*}_{0} + F^1_{-} F^{1*}_{0})}{
    |F^1_{+}|^2+|F^1_{0}|^2+|F^1_{-}|^2}\,, \\
    \eta_{3}
&=& \frac{2{\rm Re}(F^1_{+} F^{1*}_{-})}{
    |F^1_{+}|^2+|F^1_{0}|^2+|F^1_{-}|^2}\,.
\end{eqnarray}
The longitudinal $(00)$ element of the $3\times 3$ density matrix $\rho^Y$
averaged over the polar-angle distribution is given in terms of the five
polarization parameters by
\begin{eqnarray}
  \rho^Y_{00}
&=& \frac{1}{3}-\frac{1}{6} Q^{X_2} \eta_{3}
    -\frac{1}{12}\langle 3\cos^2\omega -1\rangle (2\eta_1-Q^{X_2}\eta_{3})
    \nonumber\\
&& -\frac{1}{24}\langle (3\cos^2\omega-1)(3\cos^2\theta-1)\,\rangle Q^{X_2}
    (2-\eta_1+\eta_{3})
    \nonumber\\
&& -\frac{3}{4}\langle \cos\omega\sin\omega\cos\theta\sin\theta\rangle\,
     Q^{X_2}\eta_{2}
    \nonumber\\
&& -\frac{1}{4}\langle (3\cos^2\omega-1)\cos\theta\rangle\, P^{X_2} A_{1}
   -\frac{3}{4}\langle \cos\omega\sin\omega\sin\theta\rangle\, P^{X_2} A_{2}\,,
\end{eqnarray}
and two transverse elements of the density matrix by
\begin{eqnarray}
\rho^Y_{\pm\pm} = \frac{1}{2} (\rho^Y_T \pm \rho'^Y_T)\,,
\end{eqnarray}
with the sum  and difference, $\rho^Y$ and $\rho'^Y_T$, defined as
\begin{eqnarray}
  \rho^Y_T
&=& \frac{2}{3}+\frac{1}{6} Q^{X_2} \eta_{3}
   +\frac{1}{12}\langle 3\cos^2\omega -1\rangle (2\eta_R-Q^{X_2}\eta_{3})
    \nonumber\\
&& +\frac{1}{24}\langle (3\cos^2\omega-1)(3\cos^2\theta-1)\rangle\, Q^{X_2}
    (2-\eta_{1}+\eta_{3})
    \nonumber\\
&& +\frac{3}{4}\langle \cos\omega\sin\omega\cos\theta\sin\theta\rangle\,
     Q^{X_2}\eta_{2}
    \nonumber\\
&& +\frac{1}{4}\langle (3\cos^2\omega-1)\cos\theta\rangle\, P^{X_2} A_{1}
   +\frac{3}{4}\langle \cos\omega\sin\omega\sin\theta\rangle\, P^{X_2} A_{2}\,,\\
   \rho'^Y_T
&=& \langle \cos\omega\rangle A_{1}
   +\frac{1}{2}\langle \cos\omega\cos\theta\rangle\, P^{X_2} (2+\eta_{1})
   \nonumber\\
&& +\frac{3}{4}\langle \sin\omega\sin\theta\rangle\, P^{X_2} \eta_{2}
   \nonumber\\
&& +\frac{1}{4}\langle \cos\omega(3\cos^2\theta-1)\rangle\, Q^{X_2} A_{1}
   +\frac{3}{4}\langle \sin\omega\cos\theta\sin\theta\rangle\, Q^{X_2} A_{2}\,,
\end{eqnarray}
with the longitudinal polarization $P^{X_2}=\rho^{X_2}_{++}-\rho^{X_2}_{--}$
and the (diagonal) tensor polarization $Q^{X_2}=\rho^{X_2}_{++}
-2\rho^{X_2}_{00}+\rho^{X_2}_{--}$ of the decaying particle $X_2$,
satisfying the normalization condition ${\rm Tr}(\rho^Y)=
\rho^Y_T+\rho^Y_{00}=1$.\\

\subsection{\fbox{\boldmath{$1 \to 1 + 1$}}}
\label{subsec:1_1_1}

The process $Z\to W^\pm\rho^\mp$ might be a example of this decay category,
which is yet to be confirmed experimentally. A non-standard example is the
decay of a heavy vector boson $W^\pm_H$ into two SM gauge bosons such as
$W^\pm_H \to W^\pm Z$, appearing in the little Higgs models \cite{Han:2003wu}\\

In the $X_2$ rest frame, the helicity amplitude of this type of decay modes
can be cast into the form:
\begin{eqnarray}
  D_{\sigma_2;\sigma,\sigma_1}
= F^{1}_{\sigma\sigma_1} d^1_{\sigma_2, \sigma-\sigma_1}(\theta)\,
  e^{i\sigma_2\phi}\,,
\end{eqnarray}
where the $X_2$ helicity $\sigma_2=\pm 1, 0=\pm, 0$, the $Y$ helicity
$\sigma=\pm 1, 0=\pm, 0$ while the $X_1$ particle is spinless.\\

For notational convenience, we introduce the full sum of absolute squares of
reduces helicity amplitudes $\Sigma^1_{11}$
\begin{eqnarray}
  \Sigma^1_{11}
= |F^1_{++}|^2 + |F^1_{10}|^2+|F^1_{01}|^2+|F^1_{00}|^2+|F^1_{0-}|^2
 +|F^1_{-0}|^2 + |F^1_{--}|^2\,,
\end{eqnarray}
as well as five parity-odd polarization parameters defined as
\begin{eqnarray}
    A_{1}
&=& \frac{|F^1_{+0}|^2-|F^1_{-0}|^2+2|F^1_{0+}|^2-2|F^1_{0-}|^2}{
    \Sigma^1_{11}}\,,
    \\
    A_{2}
&=& \frac{|F^1_{++}|^2-|F^1_{--}|^2+|F^1_{+0}|^2-|F^1_{-0}|^2}{
    \Sigma^1_{11}}\,,
    \\
    A_{3}
&=& \frac{2|F^1_{++}|^2-2|F^1_{--}|^2-|F^1_{+0}|^2+|F^1_{-0}|^2}{
    \Sigma^1_{11}}\,,
    \\
    A_{4}
&=& \frac{2{\rm Re}(F^1_{++} F^{1*}_{0+} - F^1_{--} F^{1*}_{0-}
                   +F^1_{+0} F^{1*}_{00} - F^1_{-0} F^{1*}_{00})}{
    \Sigma^1_{11}}\,,
    \\
     A_{5}
&=& \frac{2{\rm Re}(F^1_{++} F^{1*}_{0+} - F^1_{--} F^{1*}_{0-}
                   -F^1_{+0} F^{1*}_{00} + F^1_{-0} F^{1*}_{00})}{
    \Sigma^1_{11}}\,,
\end{eqnarray}
and six parity-even polarization parameters
\begin{eqnarray}
    \eta_{1}
&=& \frac{|F^1_{++}|^2+|F^1_{--}|^2+|F^1_{+0}|^2+|F^1_{-0}|^2
          -2|F^1_{0+}|^2-2|F^1_{00}|^2-2|F^1_{0-}|^2}{
    \Sigma^1_{11}}\,,
    \\
    \eta_{2}
&=& \frac{|F^1_{+0}|^2-2|F^1_{00}|^2+2|F^1_{-0}|^2}{
    \Sigma^1_{11}}\,,
    \\
    \eta_{3}
&=& \frac{|F^1_{++}|^2+|F^1_{0+}|^2+|F^1_{0-}|^2+|F^1_{--}|^2}{
    \Sigma^1_{11}}\,,
    \\
    \eta_{4}
&=& \frac{2{\rm Re}(F^1_{++} F^{1*}_{0+} + F^1_{--}F^{1*}_{0-}
                  - F^1_{+0} F^{1*}_{00} - F^1_{-0} F^{1*}_{00})}{
        \Sigma^1_{11}}\,,
    \\
    \eta_{5}
&=& \frac{2{\rm Re}(F^1_{++} F^{1*}_{0+} + F^1_{--}F^{1*}_{0-}
                  + F^1_{+0} F^{1*}_{00} + F^1_{-0} F^{1*}_{00})}{
        \Sigma^1_{11}}\,,
    \\
    \eta_{6}
&=& \frac{2{\rm Re}(F^1_{+0} F^{1*}_{-0}}{
        \Sigma^1_{11}}\,.
\end{eqnarray}
The longitudinal $(00)$ element of the $3\times 3$ density matrix $\rho^Y$
averaged over the polar-angle distribution is given in terms of the six
polarization parameters by
\begin{eqnarray}
  \rho^Y_{00}
&=& \rho^Y_L = \frac{1}{3}-\frac{1}{6}Q^{X_2}\eta_6
   -\frac{1}{12} \langle 3\cos^2\omega-1\rangle (2\eta_1-Q^{X_2} \eta_{6})
   -\frac{1}{4}\langle (3\cos^2\omega -1)\cos\theta\rangle\, P^{X_2} A_1
   \nonumber\\[1mm]
&& -\frac{3}{4}\langle \sin\omega\cos\omega\sin\theta\rangle\, P^{X_2} A_4
   +\frac{1}{24}\langle(3\cos^2\omega-1)(3\cos^2\theta-1)\rangle\,
                Q^{X_2} (4\eta_3-2+\eta_2-\eta_6)
   \nonumber\\[1mm]
&& +\frac{3}{4}\langle \cos\omega\sin\omega\cos\theta\sin\theta\rangle\,
   Q^{X_2} \eta_4\,,
\end{eqnarray}
and two transverse elements of the density matrix by
\begin{eqnarray}
\rho^Y_{\pm\pm} = \frac{1}{2} (\rho^Y_T \pm \rho'^Y_T)\,,
\end{eqnarray}
with the sum $\rho^Y$ and the difference $\rho'^Y_T$ defined as
\begin{eqnarray}
  \rho^Y_T
&=& \frac{2}{3}+\frac{1}{6}Q^{X_2}\eta_6
   +\frac{1}{12} \langle 3\cos^2\omega-1\rangle (2\eta_1-Q^{X_2} \eta_{6})
   -\frac{1}{4}\langle (3\cos^2\omega -1)\cos\theta\rangle P^{X_2} A_1
   \nonumber\\
&& +\frac{3}{4}\langle \sin\omega\cos\omega\sin\theta\rangle P^{X_2} A_4
   -\frac{1}{24}\langle(3\cos^2\omega-1)(3\cos^2\theta-1)\rangle\,
                Q^{X_2} (4\eta_3-2+\eta_2-\eta_6)
   \nonumber\\
&& -\frac{3}{4}\langle \cos\omega\sin\omega\cos\theta\sin\theta\rangle\, Q^{X_2}
    \eta_4\,,
   \\
   \rho'^Y_T
&=& \langle \cos\omega\rangle A_{2}
   +\frac{1}{2}\langle \cos\omega\cos\theta\rangle P^{X_2} (2+\eta_2-2\eta_3)
   +\frac{3}{4}\langle \sin\omega\sin\theta\rangle P^{X_2} \eta_{5}
   \nonumber\\
&& -\frac{1}{4}\langle \cos\omega(3\cos^2\theta-1)\rangle\, Q^{X_2} A_3
   -\frac{3}{4}\langle \sin\omega\cos\theta\sin\theta\rangle\, Q^{X_2} A_{5}\,,
\end{eqnarray}
with the longitudinal polarization $P^{X_2}=\rho^{X_2}_{++}-\rho^{X_2}_{--}$
and the (diagonal) tensor polarization $Q^{X_2}=\rho^{X_2}_{++}
-2\rho^{X_2}_{00}+\rho^{X_2}_{--}$ of the decaying particle $X_2$,
satisfying the normalization condition ${\rm Tr}(\rho^Y)=
\rho^Y_T+\rho^Y_{00}=1$.\\

\section{Polarization estimator functions}
\label{appendix:polarization_estimator_functions}

In this appendix, we exhibit all the essential functions defining the averages
of the polar-angle correlations over the polar angle $\theta$ of the products,
which consist of trigonometric functions of $\omega$ and $\theta$ explicitly in
terms of $\beta_2$ and $\beta$. We call them {\it polarization estimator
functions}, reflecting the naming {\it polarization estimators} in
Refs.$\,$\cite{V.:2016wba,Velusamy:2018ksp}. \\

For notational convenience and for the sake of discussion let us introduce the
following combinations of two speed parameters $\beta_2$ and $\beta$
as\footnote{It is interesting to note that the $n$-th power of $\beta_\pm$ is
simply $\beta^n_{\pm} = \frac{1}{2}(\beta^n_2+\beta^n\pm |\beta^n_2-\beta^n|)$
for an arbitrary integer $n$.}
\begin{eqnarray}
&& \beta_+
  =\frac{1}{2}\left(\beta_2+\beta + |\beta_2-\beta|\right)
           = {\rm max}(\beta_2,\beta)\,,\\
&& \beta_-
  =\frac{1}{2}\left(\beta_2+\beta - |\beta_2-\beta|\right)
           = {\rm min}(\beta_2,\beta)\,,
\end{eqnarray}
and three auxiliary functions of $\beta_2$ and $\beta$ defined by
\begin{eqnarray}
    {\cal L}_1(\beta_2,\beta)
 = \frac{1}{\beta_2 \beta^2}\beta_-
    -\frac{(1-\beta^2)}{2\beta_2\beta^2}
      \ln\left(\frac{1+\beta_-}{1-\beta_-}\right)\,,
\end{eqnarray}
and
\begin{eqnarray}
    {\cal L}_2(\beta_2,\beta)
 &=& \left\{ \begin{array}{ll}
     \frac{1}{\beta_2 \beta}
     \ln\left|\frac{\beta_2+\beta}{\beta_2-\beta}\right|
    -\gamma_2\gamma\frac{(2-\beta^2_2-\beta^2)}{2\beta_2\beta}
     \ln\left|\frac{\gamma_2\beta_2+\gamma\beta}{\gamma_2\beta_2-\gamma\beta}
     \right| & \mbox{for}\ \ \beta_2\neq \beta\,, \\[3mm]
     -\frac{1}{\beta^2}\ln(1-\beta^2) & \mbox{for} \ \ \beta_2=\beta\,,
     \end{array}\right.
\label{eq:1st_definition_L_2}
\\[4mm]
     {\cal L}_3(\beta_2,\beta)
 &=& \left\{\begin{array}{ll}
     \frac{(4-\beta^2_2-\beta^2)}{2\beta_2\beta}\,
           \ln\left|\frac{\beta_2+\beta}{\beta_2-\beta}\right|
    -\gamma_2\gamma \frac{(4-3\beta^2_2-3\beta^2+2\beta^2_2\beta^2)}{2\beta_2\beta}
     \ln\left|\frac{\gamma_2\beta_2+\gamma\beta}{\gamma_2\beta_2-\gamma\beta}\right|
     & \mbox{for}\ \ \beta_2\neq \beta\,, \\[3mm]
      -\frac{(2-\beta^2)}{\beta^2}\ln(1-\beta^2) & \mbox{for} \ \ \beta_2=\beta
      \,,
      \end{array}\right.
\label{eq:1st_definition_L_3}
\end{eqnarray}
with the $X_2$ and $Y$ boost factors $\gamma_2=1/\sqrt{1-\beta^2_2}$ and
$\gamma=1/\sqrt{1-\beta^2}$.  Folding these functions with proper ratios of
polynomial functions enable us to express all the polarization estimator
functions in terms of $\beta_2$ and $\beta$.\\

In order to avoid the apparently-looking singular structure in ${\cal
L}_{2,3}(\beta_2,\beta)$  with $\beta_2=\beta$ in
Eqs.$\,$(\ref{eq:1st_definition_L_3}) and (\ref{eq:1st_definition_L_3}), it is
worthwhile to reexpress the functions in a good singular-free form as
\begin{eqnarray}
  {\cal L}_2(\beta_2,\beta)
&=& {\cal L}_+(\beta_2,\beta)-{\cal L}_-(\beta_2,\beta)\,, \\
  {\cal L}_3(\beta_2,\beta)
&=& \frac{1}{\gamma_2\gamma}
    \left[(\gamma_2\gamma+1)\, {\cal L}_+(\beta_2,\beta)
         -(\gamma_2\gamma-1)\, {\cal L}_-(\beta_2,\beta)\right]\,,
\end{eqnarray}
in terms of the following two logarithmic functions:
\begin{eqnarray}
  {\cal L}_\pm (\beta_2,\beta)
\,=\, \frac{(\gamma_2\pm\gamma)^2}{4\gamma_2\beta_2\gamma\beta}
  \ln\left(\frac{\gamma_2\gamma+\gamma_2\beta_2\gamma\beta\pm 1}{
                 \gamma_2\gamma-\gamma_2\beta_2\gamma\beta\pm 1}\right)\,.
\end{eqnarray}
For $\beta_2=\beta$, we have a compact expression of ${\cal
L}_+(\beta,\beta)=-\ln(1-\beta^2)/\beta^2$ and ${\cal L}_-(\beta,\beta)=0$ with
the limit ${\cal L}_+(0,0)=1$, free from any apparent
singularities.\\

\begin{table}[t]
\centering
\begin{tabular}{|c||c|c|}
\hline
 & & \\[-2mm]
  Polarization estimator functions
& $\beta_2\to 1$
& $\beta_2=0$ or $\beta=1$
\\[2mm] \hline\hline
 & & \\[-3mm]
  $\langle \cos\omega\rangle$
& ${\cal L}_1(1,\beta)$
& $1$
\\[1mm] \hline
 & & \\[-3mm]
  $\langle \cos\omega\cos\theta\rangle$
& $ 1-{\cal L}_1(\beta,\beta)$
& $0$
\\[1mm] \hline
 & & \\[-3mm]
  $\langle \cos\omega \cos^2\theta\rangle $
& $ \frac{1}{\beta}\, \left[{\cal L}_1(\beta,\beta)-2/3\right]$
& $\frac{1}{3}$
\\[1mm] \hline
 & & \\[-3mm]
  $\langle \sin\omega\sin\theta\rangle$
& $ \sqrt{1-\beta^2}\,\, {\cal L}_1(\beta,\beta)$
& $0$
\\[1mm] \hline
 & & \\[-3mm]
  $\langle \sin\omega\cos\theta\sin\theta\rangle$
& $ \frac{\sqrt{1-\beta^2}}{\beta}\left[2/3-{\cal L}_1(\beta,\beta)\right]$
& $0$
\\[1mm] \hline
 & & \\[-3mm]
  $\langle \cos^2\omega\rangle$
& $2 {\cal L}_1(\beta,\beta)-1$
& $1$
\\[1mm] \hline
 & & \\[-3mm]
  $\langle \cos^2\omega \cos\theta\rangle $
& $ \frac{1}{\beta}\left[2-(3-\beta^2)\, {\cal L}_1(\beta,\beta)\right]$
& $0$
\\[1mm] \hline
 & & \\[-3mm]
  $\langle \cos^2\omega \cos^2\theta\rangle $
& $ \frac{1}{\beta^2}\left[\beta^2-8/3
                           +2(2-\beta^2)\, {\cal L}_1(\beta,\beta)\right]$
& $\frac{1}{3}$
\\[1mm] \hline
 & & \\[-3mm]
  $\langle \cos\omega\sin\omega\sin\theta\rangle $
& $ \frac{\sqrt{1-\beta^2}}{\beta}
    \left[3\, {\cal L}_1(\beta,\beta)-2\right]$
& $0$
\\[1mm] \hline
 & & \\[-3mm]
  $\langle \cos\omega\sin\omega\cos\theta\sin\theta\rangle $
& $ \frac{\sqrt{1-\beta^2}}{\beta^2}\left[8/3-(4-\beta^2)\, {\cal L}_1(\beta,\beta)\right]$
& $0$
\\[1mm] \hline
\end{tabular}
\vskip 0.5cm
\caption{\it Asymptotic expressions of ten polarization estimators
             to be valid when $\beta_2\to 1$, i.e. the decaying particle
             $X_2$ is highly relativistic and so greatly energetic. In addition,
             the last column shows the trivial values for $\beta_2=0$ or $\beta=1$ for which
             no Wick helicity rotation is developed.
          }
\label{tab:polarization_estimators_limits}
\end{table}

As noted before, no Wick helicity rotation is developed when $\beta_2=0$ or
$\beta=1$, i.e. $\omega =0$, leading to trivial values of the polarization
estimators.\footnote{It is unnecessary to consider the limit of $\beta=0$ as
the process $X_2\to Y X_1$ will not occur due to the vanishing phase space for
the production of $Y$ and $X_1$ at rest.} In contrast, the polarization
estimator functions have their non-trivial limits as
$\beta_2\to 1$.\\

It is a trivial observation that there is no Wick helicity rotation, if the
particle $Y$ is spinless, i.e. $j=0$. On the other hand, if the decaying
particle $X_2$ is spinless with $j_2=0$, only two polarization estimators
$\langle \cos\omega\rangle$ and $\langle\cos^2\omega\rangle$ appear in the
decay $X_2\to Y X_1$ for $j=1/2$ and $j=1$, as the decay angular distribution
in the $X_2$ rest frame is isotropic, i.e. a constant. The former estimator
function  $\langle \cos \omega\rangle$ is involved in the final-state mode
$\|\frac{1}{2}\frac{1}{2}\|$ and/or mode $\|11\|$, if parity is violated in the
decay, and the latter estimator function $\langle \cos^2\omega\rangle$ appears
in the final-state modes $\|10\|$ and $\|11\|$ with a spin-1 $Y$. Explicitly,
they can be written in terms of the functions ${\cal L}_{1,2,3}$ as
\begin{eqnarray}
    \langle \cos\omega \rangle
&=& {\cal L}_1(\beta_2,\beta)\,,
    \\
    \langle \cos^2\omega \rangle
&=& \frac{1}{\beta^2}\left[1-(1-\beta^2)\,{\cal L}_2(\beta_2,\beta)\right]\,,
\end{eqnarray}
where $\beta_2$ and $\beta$ are the speeds of $X_2$ in the LAB and $Y$ in the
$X_2$ rest frame, respectively. Their asymptotic expressions in the limit
$\beta_2\to 1$ are listed
in the second and third rows of Table~\ref{tab:polarization_estimators_limits}.\\

\begin{figure}[htb]
\centering
\includegraphics[width=15cm, height=12.5cm]{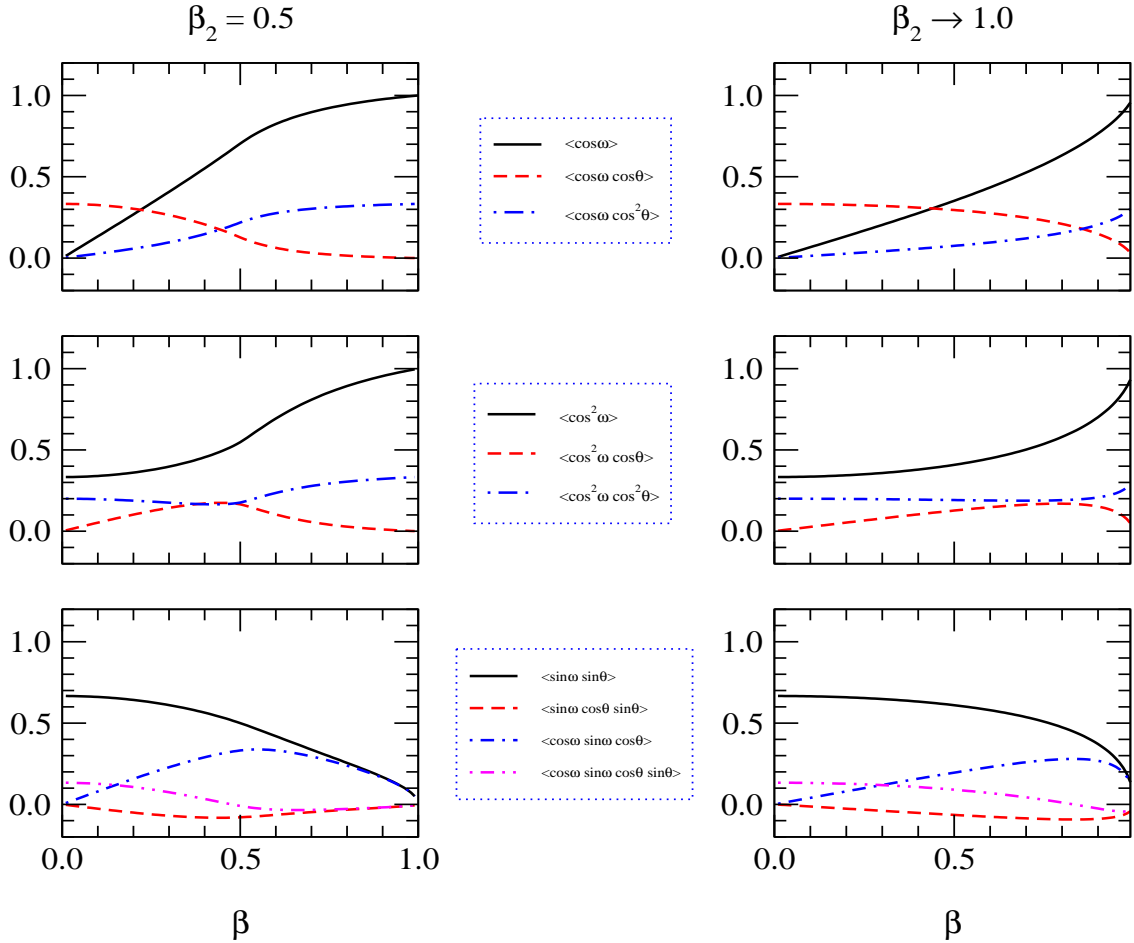}
\caption{\it The $\beta$ dependence of ten polarization estimators for fixed
             $\beta_2=0.5$ (left) and as $\beta_2\to 1.0$ (right). Note that there
             exists a rather abrupt slope change near $\beta=\beta_2$ as indicated
             clearly by the lines on the left panel.
             }
\label{fig:polarization_estimator_0.5_1.0_beta}
\end{figure}
\begin{figure}[htb]
\centering
\includegraphics[width=15cm, height=12.5cm]{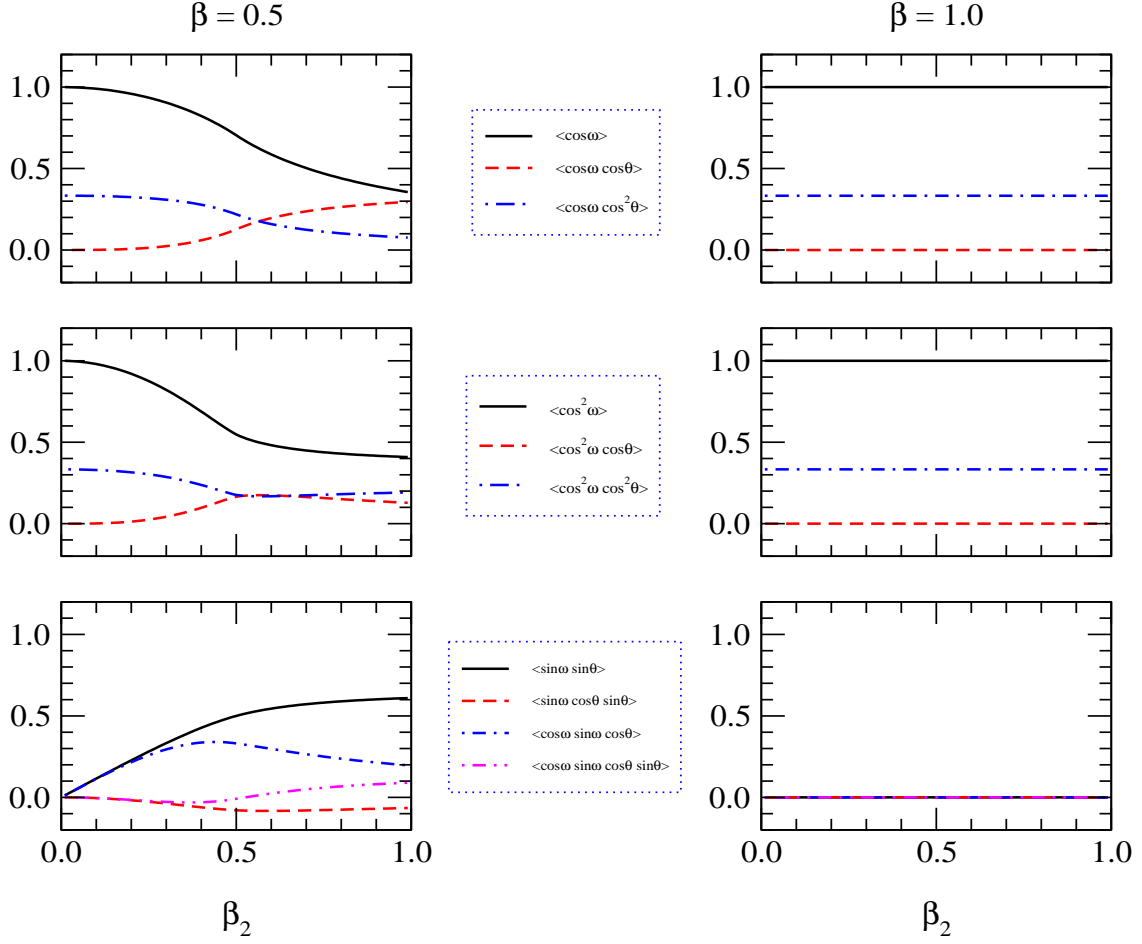}
\caption{\it The $\beta_2$ dependence of ten polarization estimators for
             $\beta=0.5$ (left) and $\beta=1.0$ (right). As mentioned in the main
             text, there is no Wick helicity rotation for $\beta=1.0$ with a massless
             $Y$, reflected clearly by the fact that all the polarization estimators
             are constant in $\beta_2$.
             }
\label{fig:polarization_estimator_0.5_1.0_beta_2}
\end{figure}

If the particle $X_2$ of non-zero spin carries non-zero polarization in a
production process of $X_2$ production, non-trivial Wick helicity rotation
effects are developed for a non-zero spin $j$ of the particle $Y$. Besides two
estimators $\langle \cos\omega\rangle$ and $\langle \cos^2\omega\rangle$, there
are eight additional non-trivial polarization estimators, involving the sines
and cosines of not only $\omega$ but also of $\theta$ explicitly. The
non-trivial functions in $\beta_2$ and $\beta$ can be classified into, firstly,
two $\cos\omega$-involved functions
\begin{eqnarray}
&&    \langle \cos\omega\cos\theta \rangle
 = \frac{1}{2\beta_2\beta}\left[
      \frac{1}{\beta}\beta_+
    -\frac{(\beta^2_2-2\beta^2)}{\beta_2 \beta^2} \beta_-
    -(3-\beta^2_2)\, {\cal L}_1(\beta_2,\beta)\right]\,,
\\
&&    \langle \cos\omega\cos^2\theta \rangle
 =-\frac{1}{2\beta^2_2\beta^2}
   \bigg[\frac{(5-3\beta^2)}{3\beta} \beta_+
        -\frac{(5\beta^2_2-10\beta^2+3\beta^2_2\beta^2)}{3\beta_2\beta^2}\beta_-
    \nonumber\\
&& \hskip 3.5cm -(5-3\beta^2_2-\beta^2+\beta^2_2\beta^2)\,
                 {\cal L}_1(\beta_2,\beta)\bigg]\,,
\end{eqnarray}
secondly, two $\sin\omega$--involved functions expressed in terms of the
logarithmic function ${\cal L}_1(\beta_2,\beta)$ as
\begin{eqnarray}
\\
&&    \langle \sin\omega\sin\theta \rangle
=-\frac{\gamma}{2\beta_2\beta}
      \bigg[\frac{(1-\beta^2)}{\beta}\beta_+
           -\frac{(\beta^2_2-2\beta^2+\beta^2_2\beta^2)}{\beta_2\beta^2}\beta_-
             \nonumber\\
&& \hskip 3.5cm
     -(3-\beta^2_2-\beta^2-\beta^2_2\beta^2)\, {\cal L}_1(\beta_2,\beta)\bigg]\,,
\\
&&    \langle \sin\omega\cos\theta\sin\theta \rangle
=  \frac{\gamma}{2\beta^2_2\beta^2}
   \bigg[\frac{5(1-\beta^2)}{3\beta}\beta_+
   - \frac{(5\beta^2_2-10\beta^2+\beta^2_2\beta^2+4\beta^4)}{3\beta_2\beta^2}\beta_-
   \nonumber\\
&& \hskip 4.2cm
  -(5-3\beta^2_2-3\beta^2+\beta^2_2\beta^2)\, {\cal L}_1(\beta_2,\beta)\bigg]\,,
\end{eqnarray}
thirdly, two $\cos^2\omega$--involved functions expressed in terms of two
logarithmic functions ${\cal L}_{2,3}(\beta_2,\beta)$ as
\begin{eqnarray}
&&    \langle \cos^2\omega\cos\theta \rangle
= -\frac{(1-\beta^2)}{\beta_2\beta^3}
     \left[2-{\cal L}_3(\beta_2,\beta)\right]\,,
\\
&&    \langle \cos^2\omega\cos^2\theta \rangle
= \frac{1}{3\beta^2}
  +\frac{1}{\beta^2_2\gamma^2\beta^4}
     \left[4-\beta^2_2-\beta^2+(\beta^2_2+\beta^2-\beta^2_2\beta^2)\,
           {\cal L}_2(\beta_2,\beta)
          -2{\cal L}_3(\beta_2,\beta)\right]\,,
\end{eqnarray}
and finally two $(\cos\omega\sin\omega)$--involved functions expressed in terms
of the logarithmic functions ${\cal L}_{2,3}(\beta_2,\beta)$ as
\begin{eqnarray}
 && \langle \cos\omega\sin\omega\sin\theta\rangle = \frac{1}{\beta_2\gamma\beta^3}
     \left[2-\beta^2+\beta^2\, {\cal L}_2(\beta_2,\beta)
          -{\cal L}_3(\beta_2,\beta)\right]\,,
\\
&&    \langle \cos\omega\sin\omega\cos\theta\sin\theta \rangle
=  -\frac{1}{\beta^2_2\gamma\beta^4}
     \bigg[4-\beta^2_2-3\beta^2+\frac{1}{3}\beta^2_2\beta^2\nonumber\\
&& \hskip 5.cm +(\beta^2_2+\beta^2-\beta^2_2\beta^2)\,{\cal L}_2(\beta_2,\beta)
               -(2-\beta^2)\,{\cal L}_3(\beta_2,\beta)\bigg]\,,
\end{eqnarray}
where $\beta_2$ and $\beta$ are the $X_2$ speed in the LAB and the $Y$ speed in
the $X_2$ rest frame and $\gamma_2=1/\sqrt{1-\beta^2}$ and
$\gamma=1/\sqrt{1-\beta^2}$, respectively.\\

The asymptotic expressions of the polarization estimator functions when
$\beta_2\to 1$, i.e. the particle $X_2$ is highly relativistic are listed in
the second column of Table~\ref{tab:polarization_estimators_limits}. In
addition, for the sake of reference, we list the values for $\beta_2=0$ and/or
$\beta=1$ in the third column of the table that are
trivially constant because of no Wick helicity rotation in those limits.\\


\end{document}